\documentclass[12pt,preprint]{aastex}
\def\HI{{\ion{H}{1}}}
\def\HII{{\ion{H}{2}}}
\def\mum{$\mu$m}
\newcommand{\Msunpyr}{\ifmmode {\rm\,M_\odot\,yr^{-1}} \else {${\rm\,M_\odot\,yr^{-1}}$}\fi}
\newcommand{\OmM}{\ifmmode {\Omega_{\rm M}}\else $\Omega_{\rm M}$\fi}
\newcommand{\OmL}{\ifmmode {\Omega_{\Lambda}}\else $\Omega_{\Lambda}$\fi}
\newcommand{\kmps}{\ifmmode {\rm\,km~s^{-1}} \else ${\rm\,km\,s^{-1}}$\fi}

\received{2004 April 2}
\begin{document}

\title{Modelling the Pan-Spectral Energy Distribution of Starburst Galaxies: I.\\
The role of ISM pressure \& the Molecular Cloud Dissipation Timescale}

\author{Michael A. Dopita, Brent A. Groves, J\"org Fischera, \& Ralph S. Sutherland}
\affil{Research School of Astronomy \& Astrophysics,
The Australian National University, \\
Cotter Road, Weston Creek, ACT 2611, Australia}
\author{Richard J. Tuffs, Cristina C. Popescu, }
\affil{Max-Plank-Institut f\"ur Kernphysik, Saupfercheckweg 1, D-69117 Heidelberg, Germany }
\author{Lisa J. Kewley,}
\affil{Harvard-Smithsonian Center for Astrophysics, 60 Garden St., Cambridge, MA , USA}
\author{Michiel Reuland}
\affil{Sterrewacht Leiden, P.O. Box 9513, 2300 RA Leiden, The Netherlands}
\author{\& Claus Leitherer}
\affil{Space Telescope  Science Institute, 3700 San Martin Drive, Baltimore  MD21218, USA}
\email{Michael.Dopita@anu.edu.au}

%\doublespace

\begin{abstract}
   In this paper, we combine the stellar spectral synthesis code STARBURST 99, the nebular modelling
code MAPPINGS IIIq, a 1-D dynamical evolution model of \HII\ regions around massive clusters of young stars 
and a simplified model of synchrotron emissivity to produce purely theoretical self-consistent synthetic 
spectral energy distributions (SEDs) for (solar metallicity) starbursts lasting some $10^8$~years. These 
SEDs extend from the Lyman Limit to beyond 21 cm. We find that two ISM parameters control the form of 
the SED; the pressure in the diffuse phase of the ISM (or, equivalently, its density), and the molecular 
cloud dissipation timescale. In particular, the shape of the FIR (dust re-emission) bump is strongly dependent 
on the mean pressure in the star-forming or starburst galaxy. This can explain the range of FIR colors seen 
in starburst galaxies. In the case of objects of composite excitation, such diagrams potentially provide a 
means of estimating the fraction of the FIR emission that is contributed by an active nucleus. We present 
detailed SED fits to Arp~220 and NGC~6240, and we give the predicted colors for starburst galaxies derived 
from our models for the IRAS and the Spitzer Space Observatory MIPS and IRAC instruments. Our models reproduce 
the spread in observed colors of starburst galaxies. From both the SED fits and the color:color diagrams, we 
infer the presence of a population of compact and ultra-compact \HII\ regions around single OB stars or small 
OB clusters. Finally, we present absolute calibrations to convert observed fluxes into star formation rates 
in the UV (GALEX), at optical wavelengths (H$\alpha$), and in the IR (IRAS or the Spitzer Space Observatory). 
We show that 25\mum\ fluxes are particularly valuable as star formation indicators since they largely eliminate 
one of the parameters controlling the IR SED.
\end{abstract}


\section{\label{intro}Introduction}
A knowledge of the star formation rate (SFR) is fundamental if we are to understand the formation and 
evolution of galaxies. In a much-cited seminal paper \citet{Madau96} connected the star formation in the 
distant universe to that estimated from low-redshift surveys, by plotting the estimated star formation 
rate per unit co-moving volume against redshift. A wide variety of techniques contribute towards 
populating this diagram with observational points. 

At the present time an unacceptable degree of uncertainty still attaches to the Madou plot, and hence to
our overall understanding of the evolution of star formation in the universe. This is largely the result 
of the  effects of dust obscuration in and around the star-forming regions, which are particularly severe
for those techniques which depend upon optical or UV data \citep[see the review by][]{Calzetti01}.
Use of the far-IR may provide improved estimates \citep{Dwek98, Gispert00}. Nonetheless, uncertainties 
still remain, since to derive these rates, one must assume that the dust is effectively acting as a bolometer
wrapped around the star forming region, and that cool and old stars do not provide too 
much of a contribution. All of these assumptions deserve closer examination in both disk and starburst
galaxies.

A further major uncertainty relates to the fact that we do not yet have reliable and self-consistent 
theoretical spectral energy distributions (SEDs) covering the full range of the electromagnetic spectrum.
In principle, almost any part of the SED of a starburst can be used as a star formation indicator, 
provided that the appropriate bolometric correction to the absolute luminosity can be made. These bolometric
corrections critically depend on the dust absorption, and the geometry of the dust clouds, both  optically
thin and optically thick, with respect to both the ionizing stars and the older stellar population. 

In principle, provided that the initial mass function (IMF) is invariant, the intrinsic UV luminosity 
should scale directly as the star formation rate. From stellar spectral synthesis models, the star formation 
rate is given in terms of the 1500\AA\ flux by \citep{Kennicutt98,Panuzzo03}:

\begin{equation}
\left[SFR_{{\rm UV }}\over \rm {M_{\odot}~yr^{-1}}\right]
=(1.2 -1.4)\times 10^{-28}\left[ L_{{\rm UV }}\over{\rm erg~s}^{-1}~{\rm Hz}^{-1}\right] .  \label{1}
\end{equation}

In practice, the results obtained with this technique depend critically on the corrections 
needed to account for the dust in and around the star-forming galaxy. These are very uncertain; there are 
claims that a typical $z=3$ galaxy suffers a factor of 10 extinction at a wavelength of 1500\AA\ 
in the rest--frame of the galaxy \citep{Meurer99,Sawicki98}. Others argue for more modest 
corrections \citep[e.g.][]{Trager97}. In normal galaxies, \citet{Bell01} find that 
although mean extinctions are modest (about 1.4~mag at 1550\AA ), the scatter from galaxy to galaxy is very 
large. For individual galaxies at high redshift the reddening and the SFR are strongly correlated, and since  
many fainter galaxies at high redshift are missed completely, potentially large uncertainties exist in the 
derived star formation rates \citep{Adelberger00}.

A fairly direct and extensively-used technique is to measure hydrogen recombination line fluxes 
\citep[][ to name but a few]{Bell01,Gallego95,Jones01, Moorwood00,Tresse98,Pascual01,Yan99,Dopita02}.
Provided that the \HII\ region can absorb all the EUV photons produced by the central stars, this should be 
a reliable technique, since in this case the flux in any hydrogen line is simply proportional to the number 
of photons produced by the hot stars, which is in turn proportional to the birthrate of massive stars. This 
relationship has  been well calibrated at solar metallicity for the H$\alpha $ line. In units  
of M$_{\odot}$~yr$^{-1}$, the estimated star formation rate is given by 
\citep{Dopita94,Kennicutt98,Panuzzo03}:

\begin{equation}
\left[SFR_{{\rm H\alpha }} \over \rm {M_{\odot}~yr^{-1}}\right]
=(7.0-7.9)\times 10^{-42}\left[ L_{{\rm H\alpha }}\over {\rm erg.s}^{-1}\right] .  \label{2}
\end{equation}

Provided that the star-formation regions are resolved, their Balmer decrements can be used to estimate the
absorption in any foreground dust screen.  However, it is possible that some star forming regions are 
completely obscured, even at H$\alpha$. This problem can be avoided by observing at infrared wavelengths, 
in Br$\alpha ,$ for example. A further complication is that the dust content of the nebula is 
metallicity-dependent, and therefore an appreciable fraction of the hydrogen ionizing photons may be absorbed by
dust in high metallicity \HII\ regions.  This possibility, first seriously quantified by \citet{Petrosian72}, 
has been discussed by a  number of authors since \citep{Panagia74, Mezger74, Natta76, Sarazin77, Smith78, 
Shields95, Bottorff98}. Its effect has been investigated and quantified (as far as is possible by direct 
observation) in a series of recent papers by Inoue and his collaborators \citep{Inoue00, Inoue01a, Inoue01b}. 
\citet{Dopita03} has shown that this effect increases in importance in the more compact  \HII\ regions, 
and  may lead to errors as  high as $\sim 30$\% in the  estimated SFR.

In the case of dusty starburst galaxies, the total 8-1000~$\mu $m infrared luminosity, $L_{{\rm IR}},$ as 
measured in the rest frame frequency of the galaxy can be used to estimate the total SFR
\citep{Kennicutt98, Panuzzo03}:
\begin{equation}
\left[SFR_{{\rm IR }}\over \rm {M_{\odot}~yr^{-1}}\right]
=(4.5-4.6)\times 10^{-44}\left[ L_{{\rm IR }} \over {\rm erg.s}^{-1}\right] .  \label{3}
\end{equation}

In deriving the proportionality between IR flux and SFR, it is generally assumed that the dust acts as a
sort of ``bolometer", wrapped around the star formation regions, and re-emitting most, if not all, the 
incident EUV and FUV photons. In practice, this is one of the many ``geometrical" factors which render
the derivation of the SFR quite uncertain. For example, \citet{Inoue02} includes a number of these geometrical
factors in the following formula for the IR luminosity of a galaxy:

\begin{equation}
L_{{\rm IR}} = L_{{\rm Ly\alpha}} + (1-f)L_{{\rm LC}} +
\epsilon L_{{\rm UV}} + \eta L_{{\rm old}} \label{eqn4}
\end{equation}

This allows a fraction, $(1-f)$, of the ionizing flux to be absorbed by dust in the {\ion{H}{2}} region, 
assumes that all of the Lyman$\alpha $ photons are multiply scattered by the gas and ultimately absorbed by 
dust in the surrounding {\ion{H}{1}} region, that a fraction, $\epsilon $, of the non-ionizing UV photons are 
absorbed, and that a fraction $\eta $ of the radiation field produced by the old stars is also absorbed by 
dust. Photoionization models of {\ion{H}{2}} regions show that, typically, the Lyman$\alpha $ flux is of 
order 30\% of the total stellar flux absorbed by the nebula, so that globally, the dust is fairly efficient in 
capturing and re-radiating the flux originally radiated by the star in the Lyman continuum. Indeed, since most
of the FIR flux results from the absorption of both the non-ionizing and ionizing photons in the 
photodissociation regions around \HII\ regions, and since the temperature of the dust  depends critically
on the distance  of the dust from the  stars, it is clear that a clear understanding of the geometry
of the dusty \ion{H}{1} and molecular gas with respect to newly-born  stars is  critical to our understanding
of the SED.

In combination with STARBURST 99 \citep{Leitherer99} modelling, \citet{Hirashita03} have applied equation \ref{eqn4} to 
the observations of star forming galaxies in  order to estimate the empirical values for the geometrical 
factors, which provide estimates for SFRs consistent between the UV, IR and  H$\alpha$ estimators.  

Recently, several very sophisticated approaches to the determination of the SED in galaxies have been 
developed. For normal galaxies the models of \citet{Silva98} explicitly allow for the clumpy distribution
of dust, strongly spatially correlated with the UV-emitting young stellar clusters, and probably associated
with the dense molecular clouds from which the young stars have been formed. The model of \citet{Popescu00},
subsequently extended in later papers \citep{Misiriotis01,Tuffs04}, explicitly takes into account the 
three-dimensional stucture of spiral galaxies both in terms of their stellar and gas and dust distributions
to provide projected photometric properties and attenuation characteristics.

For starburst galaxies, \citet{Hirashita02} and \citet{Takeuchi03} have developed a dust re-emission model 
appropriate to low-metallicity young galaxies at high redshift. \citet{Takagi03a} and \citet{Takagi03b} 
have developed a mathematically elegant spherically  symmetric radiative transfer model which 
includes the very important temperature fluctuations of the small grain populations and emission in the PAH 
features between 3 and 30\mum. Applied to starburst systems, this model suggests that star formation rates
are positively correlated with the structural characteristics of the star forming region: regions of high
star formation rate are both compact and highly obscured at optical wavelengths.

Because of their effect on both the efficiency of the production of IR emission and on the form of
the IR emission through the dust temperature distribution, these geometrical factors largely determine 
the far-IR SED. In a pair of recent papers Dale and his co-workers have used ISO observations to
develop a (largely empirical) family of templates to fit the variety of the forms of the IR SED 
\citep{Dale01,Dale02}. This work assumes a power law distribution of dust mass with local radiation
field intensity to provide a wide range of dust temperatures. It appears to indicate that the SEDs of 
star forming galaxies could fit to a one-parameter family of curves determined essentially by the 
exponent, $\alpha$, of the power law distribution of the dust masses with the radiation field intensity, 
where the radiation field is assumed to have the color of the local ISRF. This will not be the case in 
starburst galaxies, where the radiation field will be harder than the local ISRF, and where the geometry 
of the dust with respect to the gas is determined by the evolution of \HII\ regions. Also, it may not be 
applicable even in normal galaxies, which are known through direct observation to appear as the 
superposition of different geometrical components whose dust emission has different intensities and colors
\citep{Tuffs03}. The commonly used 60/100 \mum\ color diagnostic, cannot be interpreted in terms of
a single grain emission component, since the 60\mum\ emission is seen to come mainly from warm dust 
locally heated in HII regions, the 100\mum\ emission comes both from the HII regions and the cold diffuse 
inter-arm emission.

In this paper, we attempt to provide a theoretical approach to the generation of pan-spectral energy 
distributions in the context of starburst galaxies. This provides physical insight into how the geometrical 
factors discussed above are determined largely by the temporal evolution of the \HII\ region complexes 
in starburst galaxies. We have included the dust physics and radiative transfer both in the ionized 
\HII\ plasma, and in the dense molecular clouds associated with them. In particular, we will show that the 
stochastic average pressure within the star formation region plays a major role in determining 
the FIR SED through its effect on the grain temperature, and that the rate at which the molecular 
clouds are dispersed and destroyed by the expanding \HII\ complexes is fundamental for our understanding 
of the geometrical factors which affect the global SED.

This paper is organized as follows. In Section \ref{Starmodel} we describe the STARBURST 99 modelling which 
provides the input stellar spectra, and in Section \ref{HIImodel} we uses the mechanical energy input given
by the STARBURST 99 modelling to build a simple one-dimensional temporal evolution model for the \HII\ regions 
surrounding a typical OB stellar cluster. In Section \ref{Dustmodel} we describe the dust model used in this 
model, and show how the photodissociation of PAH molecules combined with a turbulent ISM may explain the 
observed attenuation curve of starburst galaxies. The MAPPINGS IIIq \citep{Sutherland93, Dopita02b} modelling of 
the emission line, atomic continuum, and dust re-emission spectrum  of these \HII\ regions and their 
surrounding photodissociation regions (PDRs) and the generation from these models of the global starburst SEDs 
is described in Section \ref{Staburst}. In Section \ref{SED} we move on 
to discuss the parameters which have a controlling influence of the form of the theoretical SEDs, and go on to 
compare the models with observations in Section \ref{Obs}. Finally, in Section \ref{SFRs} we use the model to 
provide a theoretical re-calibration of the various star formation indicators. In the conclusion, we enumerate 
the main results of the paper, and we emphasize the limitations of the models.

All the models presented in this paper have a solar abundance set. In Paper II of this series, we will 
investigate the effect of chemical abundance on the form of the SEDs.

\section{Stellar Spectral Synthesis Modelling}\label{Starmodel}

Starburst activity was first linked to galaxy interactions decades ago. In 1967, \citet{Sersic67} found 
that galaxies harbouring hotspots all had extreme blue colours in their centers, a property shared by 
the majority of interacting or peculiar galaxies in the \citet{Vorontsov59} and Arp catalogues. 
An excess of radio emission was found in interacting galaxies, strengthening further the link between 
star formation and interaction \citep{Sulentic76, Condon78, Hummel80}.  \citet{Larson78} first suggested
that tidal forces in interacting galaxies could trigger bursts of star formation.  Since then, 
numerous studies have produced evidence for interaction-induced starburst activity 
\citep[{\it eg.},][]{Petrosian78, Condon82,Keel85,Madore86,Kennicutt87,Bushouse87,Hummel90,Donzelli97,
Barton00,Petrosian02,Lambas03}.

The typical duration of such interactions is a dynamical timescale; comparable to the orbital timescale 
of stars in the gravitational well of the system, $\sim 10^8$~years. The details of the star formation
triggered by such interactions depend on the many parameters of the interaction such as the impact 
parameter, the relative velocities, the relative orientations of the spin axes, and the relative masses of the
systems. Simulation of galactic mergers \citep{Barnes96} suggests that the most violent star formation 
tends to be triggered in two separate episodes; during the initial collision of the systems and later, when
tidal tail debris rains down into the nuclear regions of the merged systems. Enhanced star formation can 
occur throughout the interaction.

Given that each case will be different, for simplicity in what follows, we have assumed that in our model 
starbursts, stars are being created continuously as separated $10^4$~M$_{\odot}$ clusters scattered
through the same volume, so as to give a mean star formation rate of $1$~M$_{\odot}$~yr$^{-1}$. Thus, in the 
total lifetime of the starburst, $10^8$~years, some $10^4$ such clusters will be formed.

Clusters of young stars are dust-enshrouded for the first $\sim 1-2$~Myr, and produce the majority of their 
ionizing UV photons within the first $\sim 4-6$~Myr of their lifetime. Thus star formation indicators based
upon nebular emission lines such as H$\alpha$ or [\ion{O}{2}], or else based upon dust re-emission processes 
such as the 60 \mum\ continuum flux are strongly biased towards measuring recent star formation rates.  By
contrast, features which more directly measure the older population or the total luminosity, such as the 
near-IR continuum or the PAH emission features measure the star formation rates integrated over somewhat longer
periods. Assumptions about the total duration of the starburst do not greatly affect the calibration of the 
star formation rates based on either the UV, emission line, or FIR SED - the total luminosity of starburst
increases by only by 10\% between $5\times 10^7$~years and $10^8$~years according to the STARBURST 99 models 
\citep{Leitherer99}.

We have used the most recent version of the STARBURST 99 code to generate the theoretical 
stellar SEDs of:
\begin{itemize}
\item {Instantaneous starbursts, sampled at 1 Myr age intervals from
0 Myr up to 10 Myr and}
\item{Continuous star formation at the rate of
$1$~M$_{\odot}$~yr$^{-1}$ up to a maximum age of $10^8$~years.}

\end{itemize}

These models assume solar metallicity, a standard Salpeter initial mass function, and upper and lower mass 
cutoffs of 120~M$_{\odot}$\ and 1~M$_{\odot}$\ respectively on the zero-age main sequence.  The stellar
atmosphere models for stars with plane parallel atmospheres are based on the \citet{Kurucz92} models as 
compiled by \citet{Lejeune97}.  The fully line-blanketed Wolf-Rayet atmosphere models of \citet{Hillier98} 
and the non-LTE O-star atmospheres of \citet{Pauldrach01} have been incorporated into STARBURST 99 as 
described in \citet{Smith02}. We use the full suite of atmosphere models to provide the most up-to-date 
stellar radiation field possible. All STARBURST 99 models were re-binned to the lower spectral resolution input 
energy bins required by MAPPINGS IIIq.

The instantaneous burst models were used as the input spectra for the \HII\ region nebular gas and dust 
modelling, while the continuous star formation models form a theoretical global SED for the older stars with 
ages between 11 and 100 Myr.  The contribution of the older stars was distributed between the \HII\ regions
of different ages in proportion to their volume, and added to the stellar SEDs of the young cluster which principally
excite the \HII\ regions. Thus our global SED is composed of the sum of all the contributions from the stars 
and their surrounding \HII\ regions which provide an average star formation rate of $1$~M$_{\odot}$~yr$^{-1}$ 
for a total time of  $10^8$~years, finely sampled in time at 1~Myr time intervals between 0 and 10~Myr.

\section{The Dynamical Evolution of \HII\ Regions}\label{HIImodel}

It is very easy to understand why the temporal evolution of the \HII\ regions surrounding young clusters
of stars might be quite important in influencing the SED of starburst galaxies. First, the inner radius, 
$R_{\rm in}$, of the ionized shell and the pressure in the \HII\ region together determine the 
ionization parameter, ${\cal U} =F_*/4\pi c R_{\rm in}^2 n_{\rm HII}$, where $F_*$ is the flux of ionizing 
photons from cluster stars, c is the speed of light and $n_{\rm HII}$ is the particle density in the 
\HII\ region.  As is well understood from photoionization theory, see \citet{Dopandsuth03}, the 
emission line spectrum of an  \HII\ region is determined primarily by the effective temperature of the 
cluster stars (itself a function of time) and by the ionization parameter, which is determined by the 
radius~:~time, and the radius~:~pressure  relationships. These  relationships need to be derived from  
models of the dynamical evolution of the \HII\ regions, and are determined by the deposition of mechanical 
energy and momentum from the stellar winds and by supernovae. 

The FIR SED is also influenced by these parameters, because, provided the non-ionizing UV is
absorbed on a scale less than the radius of the HII region, the outer radius of the \HII\ region largely
determines the characteristic dust temperature for a given stellar luminosity, and the IR PAH features are
generated mostly in the photodissociation region surrounding the \HII\ region. 

In essence, then, the evolution of  the \HII\ regions determines the geometry of the gas and the dust 
with respect to the hot stars, and these geometrical effects determine various aspects of the SED. These 
effects have not hitherto been taken into account in any self-consistent way in the computation of SEDs. The
most sophisticated treatment so far  can  be found in the ``semi-analytic" framework of 
\citet{Granato00,Bressan02} and \citet{Panuzzo03}. In this, the  young  stars are treated as escaping from
a  dense molecular cloud environment on a characteristic time scale which is environment dependent. 
However, such a treatment cannot properly take into account either the change of ionization parameter within the 
\HII\ regions, nor the change in the mean dust temperature with time.

The theory of the size distribution of \HII\ regions has been presented by \citet{Oey97} and \citet{Oey98}.
According to these papers, the \HII\ regions expand and evolve as mass-loss and supernova driven bubbles
for as long as their internal pressure exceeds the ambient pressure in the ISM. When this condition is no longer
met, the \HII\ region is assumed to ``stall". Of course, on reaching the stall condition, the \HII\ region 
does not abruptly stop expanding, but a momentum-conserving expansion continues until interstellar turbulence 
destroys the integrity of swept-up shell. The time taken to reach the stall radius turns out to be
proportional to the stall radius, and the mass of the OB star cluster is important in determining the stall
radius; low mass cluster \HII\ regions stall early and at small radius.

This theory emphasizes the important role of the pressure in the ISM in determining the radius of
an individual \HII\ region, and its evolution with time. In particular, with higher ISM pressure, the 
\HII\ region stalls at smaller radius, and therefore the mean dust temperature in the atomic and molecular
shell around the \HII\ region should be higher in high-pressure environments. We would therefore expect 
the FIR SED to peak at shorter wavelengths in high-pressure starburst regions. It is this correlation that
we seek to investigate in this paper.

According to the mass-loss driven bubble theory of \citet{Weaver77}, the radius, $R$, and the internal pressure, 
$P$, of an \HII\ region are given by (\emph{c.f.} equations 24 and 25 of \citet{Oey97});

\begin{equation}
R = \left(250 \over 308\pi\right)^{1/5}L_{\rm mech}^{1/5}\rho_0^{-1/5}t^{3/5} \label{eqn5}
\end{equation}
and
\begin{equation}
P = {7 \over (3850\pi)^{2/5}}L_{\rm mech}^{2/5}\rho_0^{3/5}t^{-4/5}. \label{eqn6}
\end{equation}
Here, $L_{\rm mech}$ is the mechanical luminosity of the mass-loss from the central stars, $\rho_0$ is the density 
of the ambient medium and $t$ is the time. The particle density is given in terms of the ambient density by
$n=\rho_0/\mu m_H$, and the ambient pressure $P_0=nkT_0$. Eliminating $t$ between equation 
\ref{eqn5} and \ref{eqn6}, and setting $P =P_0$ we obtain the following conditions for stall:
\begin{eqnarray}
P = {7 \over (3850\pi)^{2/5}}\left(250 \over 308\pi\right)^{4/15}{\left(L_{\rm mech} \over \mu m_Hn\right)}^{2/3}
{\mu m_Hn \over R^{4/3}} \\ 
=nkT_0 = n_{\rm HII}kT_{\rm HII}. \label{eqn7}
\end{eqnarray}

These equations imply that, for any given pressure in the ISM, all stalled \HII\ regions have a common
ratio $L_{\rm mech}/n_{\rm HII}R^2$. Thus to the extent that $F_* \propto L_{\rm mech}$, where $F_*$ is the
flux of ionizing photons from the central stars, then all stalled \HII\ regions will be characterized 
by a common value of the ionization parameter, ${\cal U} =F_*/4\pi c R_{\rm in}^2 n_{\rm HII}$. In fact, the 
STARBURST 99 models show that the ratio of mechanical luminosity to ionizing photons may vary by as much as 
a factor of ten during the first $10^7$ years of evolution. Nonetheless, most observed extragalactic \HII\ 
regions show a fairly limited range in their ionization parameters \citep{Kewley02}, and most \HII\ regions
should be in the stall condition \citep{Oey97}, so that the above argument may well represent the basis of a 
theoretical explanation of this curious observational result.

Although we have emphasized the role of pressure, equation \ref{eqn7} shows that the stall condition 
might be better cast as a relationship between radius and ambient density, since it is the ambient density
that determines the expansion history of the bubble. The effective ambient density is set by the density
of the phase of the ISM which has the largest volume filling factor. We have assumed that, in starburst
galaxies, this phase is a warm ionized medium (WIM) or warm neutral medium (WNM) heated by the photons and 
mechanical energy of the starburst itself. Since both of these phases have a temperature $T \sim10^4$K, they
also have similar densities for a given pressure.

Whether or not the \HII\ regions are in the stall condition, the observed ionization parameter provides
an important constraint on the internal pressure, and, more importantly, on the radiation field density 
incident on photodissociation region (PDR) around the \HII\ region. This determines the characteristic 
dust temperature, and hence the wavelength of the peak in the FIR dust re-emission feature. 

\HII\ regions in the stalled condition are found preferentially in regions of higher ISM pressure or density,
and around lower mass clusters. Here, we consider the variation in pressure, but we have not taken into account
the mass distribution of the central clusters, using instead a `representative' cluster of mass $10^4M_\odot$.
Clearly, the cluster mass distribution is an important parameter, the investigation of which is deferred for
a future paper.

In the stall condition the expansion of the bubble does not cease entirely. By the time the stall radius is
reached there is already a massive shell of swept-up gas, and this carries a  substantial momentum. In our
evolutionary calculations, we have assumed that the evolution of the bubble subsequent to the stall condition 
being reached continues according to the momentum conserving solution for the swept-up shell:

\begin{equation}
R=\left[ {3M_{\rm stall}v_{\rm stall}} \over{\pi \rho_0} \right]^{1/4}t^{1/4} \label{eqn9}
\end{equation}
where the mass and velocity of the swept up shell at the stall radius are $M_{\rm stall}$ and $v_{\rm stall}$, 
respectively.

\subsection{Bubbles with Time Dependent Luminosity}
We have  used the stellar wind and supernova power output for a $10^4M_\odot$ cluster as predicted by the 
STARBURST 99 code to  make a  Runge-Kutta integration of the equation of  motion of the \HII\ shell. 
The internal pressure is calculated from equation \ref{eqn7}, with  the  instantaneous power output  being 
smoothed over  a period of 1Myr to eliminate  short time-scale pressure fluctuations.

For a time dependent wind + supernova mechanical luminosity $L_{\rm mech}(t)$ and an ambient medium density 
of $\rho_0$, the equations of conservation of mass, momentum and energy for the mass $M$ of the swept-up shell 
for a stellar wind bubble are; 

\begin{eqnarray}
\frac{d}{dt}[ M ] & =  & 4 \pi R^2  \rho_0 \dot{R} , \\
\frac{d}{dt}[ M \dot{R} ] & =  & 4 \pi P R^2 , \\
\frac{d}{dt}[ P R^3] & =  & \frac{L_{\rm mech}(t)}{2 \pi} - 2 P R^2 \dot{R} , 
\end{eqnarray}
where $R$ is the outer shell radius.  Here $P$ is the bubble pressure, and  $P >> P_0$. By assuming that the 
bubble pressure is much greater than the ambient pressure, we have an energy driven bubble. This will be true 
until shortly before the stall radius is reached.

Assuming that $\gamma = 5/3$ defines the equation of state of the bubble gas, we can combine these into a single 
differential equation, which is numerically integrated to allow for an arbitrary, time dependent, 
mechanical luminosity $L_w$,

\begin{equation}
 \frac{d}{dt}[R \frac{d}{dt}(R^3 \dot{R})] +
\frac{9}{2} R^2 \dot{R}^3 = \frac{3 L_{\rm mech}(t)}{2 \pi \rho_0}. \label{eqn13}
\end{equation}  
This equation was then integrated using a standard Runge-Kutta 4th order integration with an adaptive step 
size, using the STARBURST 99 generated tabulated input mechanical energy luminosity for our cluster as a function 
of time until the stall condition was reached, after which the momentum-conserving solution was adopted 
(equation \ref{eqn9}).

This simple integration does not yield results which are in accord with observations. The bubble expands to 
too large a diameter without reaching the stall condition, and the internal pressure is too high. As a result,
the ionization parameter in the ionized shell is about  a factor of ten below the observed values for 
extragalactic \HII\ regions as determined by \citet{Kewley02} at all times during the lifetime of the bubble.

This problem can be rectified by assuming that the mechanical power output of the young stars is much less 
effective than the simple $1-D$ hydrodynamical models would suggest in terms of its efficiency in inflating 
the ionized bubble. Empirically, we determined that the effective power input of the stars would have to be 
decreased by a factor of ten to ensure that the ionization parameter in the  \HII\ region matches the observed 
range of  $3 \times 10^{-4} \leq \rm{U} \leq 3 \times 10^{-3}$ \citep{Kewley02}. This is  in part due to
the decrease in the radius of the bubble, and in part due to the lower internal pressure. 

Interestingly enough, similar discrepancies between observation an the  simple mass-loss bubble theory have 
emerged in analyses of the dynamics of bubbles and  of their stellar content \citep{Garcia95,Oey96,Naze01}.
In these studies discrepancies of between one and two  orders of magnitude have  been noted. In all cases, 
the energy flux in the stellar winds predicted by the analysis of the stellar content exceeds the energy flux
inferred from the dynamics. These discrepancies probably arise out of the limitations  of the $1-D$ 
hydrodynamic models. Such models cannot take into account  a number of effects which lead to a decreased 
efficiency in inflating the mass-loss bubble. These include enhanced radiative losses in the shocked stellar 
wind plasma through operation of the Vishniac instability, blowouts caused by the breakup of the atomic and  
molecular shell  surrounding the \HII\ region, and `poisoning' of the hot plasma through photo-ablation 
and entrainment of fragments within the hot shell.

In the light of this, we  have modelled our $10^4M_\odot$ clusters as if they only had the mechanical energy  
luminosity  of a $10^3M_\odot$ cluster. This ensures that our model \HII\ regions have ionization parameters
which accord with observations, and this will also ensure that the dust grains in the PDRs of these bubbles will 
be illuminated by a radiation field of the correct strength, so that the grains will have the correct 
temperatures.

\begin{figure}
  \caption{\label{fig1}
    The time variation of the  radius  of the \HII\ bubble as a function of the pressure in the ISM}
\end{figure}

\begin{figure}
  \caption{\label{fig2}
    The time variation of the  interior pressure  of the \HII\ bubble as a function of the 
    pressure in the ISM}
\end{figure}

In figure \ref{fig1} we plot the computed time evolution of the radius of the bubble  as a function of the 
mean pressure $P/ k$ (measured in units of cm$^{-3}$K). The undisturbed gas outside the bubble is assumed
to be part of the warm ionized medium, so that $P/ k  = 10^4$ corresponds to an atomic density of 
$\sim 1.0~\rm {cm}^{-3}$. In figure \ref{fig2} we plot the corresponding evolution of the internal pressure
in the  bubble, which can be identified as the pressure in the \HII\ region. Note the peak in the pressure
which appears near log(t)$ \sim 6.5~$yr, which corresponds to the onset of the energy input by 
Type II supernovae.

For $P/k  <  10^6$ the bubble does not stall until quite late in its evolution, but for the highest 
pressures the stall condition is reached very early on. This ensures a large range in the final radii. 
Furthermore, for all the models,the pressure inside the bubble is quite strongly correlated with the 
external pressure, for times greater than about 1Myr and up to the time at which the Type II supernova 
energy input switches off. This suggests that a measurement of the density in the ionized plasma using a 
density-sensitive ratio such as the [\ion{S}{2}] $\lambda \lambda 6717/6731$\AA~ ratio could be used to 
help estimate the mean ambient ISM pressure in the star-forming region.

During the evolution of the \HII\ bubble, the distribution of the radiation field within the bubble can 
be modified by the dynamical evolution of the central cluster, and indeed, by the turbulent break-up of the 
shell of gas around the \HII\ region. For, example, at $10^7$yr the mean expansion speed of 
the \HII\ bubble is 0.6, 5.7 and 17.3~kms$^{-1}$ for $P_0/k = 10^8, 10^6$ and $10^4$~cm$^{-3}$K,
respectively. Thus, at the higher pressures, the velocity dispersion of the stars in the central star 
cluster is likely to rival or exceed the expansion of the \HII\ region, leading to a breakdown in the 
assumption that the illumination of the PDR at the edge of the bubble is from a single source at the center 
of the bubble. Under such circumstances, the rate of expansion of the bubble may be determined
by the velocity dispersion of the stars rather than the hydrodynamical effects, with the ``bubble" in fact 
being a cluster of smaller \HII\ regions from individual stars, rather than a single cavity. 
In this case, the picture might be of a molecular cloud being excavated from within through the growth of 
independent HII regions inflated by single stars (or by small numbers of stars), eventually creating a porous 
parent molecular cloud.

All of this serves to highlight the fact that the time has come to abandon the simple 1-D evolutionary
hydrodynamic model in favour of an approach which properly takes into account both the dynamics of the 
central cluster and the hydrodynamics of both the ionized and the molecular gas. This treatment is beyond 
the scope of this paper.

\section{Treatment of Dust and PAH Physics}\label{Dustmodel}

\subsection{\label{Molecule}The PAH Molecules}
The numerous emission features in the $3-30$\mum\ region, formerly referred to as the Unidentified Infrared 
Emission Bands or UIBs were first identified as due to polycyclic aromatic hydrocarbons (PAHs) by \cite{Leger84}.
Since that time, our understanding of these features has become rather sophisticated, as more and more
observational constraints have been produced, principally by the ISO mission \citep{Peeters02}. At the time of
writing, the Spitzer Space Observatory is providing stunning new images in the PAH bands.
Clearly, any serious attempt to model SEDs must now include a good deal of PAH physics, and in the following
subsections we describe the treatment of PAHs within the mappings code.

In modelling the PAHs within photoionized regions we have opted for simplicity over an exact treatment, especially 
as the constraints on both the physics and the species of PAHs remain so poor. In the MAPPINGS IIIq code we have
simply assumed that all the PAHs can be represented by a single representative type, coronene (C$_{24}$H$_{12}$). 
Not only has coronene been observed within the ISM, but it is also pericondensed, reasonably stable and can 
adequately match the observational constraints in the UV and IR. Though perhaps slightly smaller than the 
average ISM PAH \citep{Allamandola99}, it has the great advantage that there exists a large amount of 
laboratory data for it, hence any modelling done with these data can be considered reasonably accurate from
a purely physical viewpoint. 

The spectroscopic properties (emission and absorption) of a PAH molecule depends upon the number of carbon 
atoms (N$_\mathrm{C}$), the hydrogenation (H per C) and the charge state of the molecule. In addition to these, 
the optical properties are affected by the shape of the PAH (cata- or peri- condensed) and the addition of end 
groups (like methyl -CH$_3$). In the modelling, we assume a pericondensed structure with no additional end groups.
The PAH abundance within the code is determined by the amount of Carbon depleted onto PAH molecules. 
This was chosen to provide a good match to the absolute strength of the PAH emission features in the infrared. 

The use of a single species of PAH makes the determination of PAH survival, optical properties and charge 
much easier to determine. The absorption, emission and photoelectric processes on a single size and
shape are also much more readily calculated. \citet{Li01}, using all this available data have constructed 
absorption cross sections for both neutral and ionized ``astronomical'' PAHs from the FUV to FIR. They did this 
for a range in N$_\mathrm{C}$, with the largest of PAHs blended in with the optical properties of graphite
grains.

Within MAPPINGS IIIq, we have used their data \footnote{Available at http://www.astro.princeton.edu/$\sim$draine
/dust/dust.diel.html} to represent the PAH optical properties. The PAH optical data of \citet{Li01} is defined 
in terms of a PAH ``radius'', such that there is a continuum in data between PAH and graphite grains. This PAH 
radius connects directly with N$_\mathrm{C}$, with $a=10(\mathrm{N}_\mathrm{C}/468)^{1/3}$\AA. This gives for coronene an effective radius of $a=3.7$\AA\ and surface area of $173$\AA$^2$. These values, and the
corresponding \citet{Li01} data for this size, are used within the code to define the effective PAH molecule.
We use both the neutral and ionized PAH optical data for the code.The fraction contributed to the total 
PAH opacity is determined by the charge state of the model PAH, which is determined by the
balance of electron sticking and photoelectric ejection.

\subsection{The PAH Emission Spectrum}

The emission spectrum of PAHs is determined by the natural modes of vibration, bending and other deformation of 
the carbon skeleton, modified by the effect of the dangling groups and the electric charge state of the molecule. 
The principal modes and their characteristic wavelengths are as follows: 
\begin{itemize}
\item{3.29\mum\ Aromatic C-H stretch ($v=1-0$);}
\item{3.40\mum\ Aliphatic C-H stretch (the $v=1-0$ anti-symmetric mode);}
\item{3.51\mum\ Aliphatic C-H stretch (the $v=1-0$ symmetric mode); }
\item{6.2\mum\ and 7.7-7.9\mum C-C skeletal deformations;}
\item{8.6\mum\ C-H in-plane bend; }
\item{11.2\mum\ C-H out-of-plane bend (solo mode);}
\item{11.9\mum\ C-H out-of-plane bend (duo mode);}
\item{12.7\mum\ C-H out-of-plane bend (tri mode);}
\item{15-20\mum\ C-C-C bending modes \citep{VanKerckhoven00}.} 
\end{itemize}
These emission features
can be adequately represented by Lorentzian fits \citep{Boulanger98} of the form:

\begin{equation}
F\left( x\right) =\frac{\left( A/\pi \sigma \right) }{\left[ 1+\left(
x-x_{0}\right) ^{2}/\sigma ^{2}\right] } \label{eqn14}
\end{equation}
in which $x=1/\lambda $ cm$^{-1}$, the central wavenumber of the feature is $x_{0}$, the 
$\mathrm{FWHM}=2\sigma ,$ and the area under the curve is $A.$ 

Two model fits were made, the first to the data of \citet{Boulanger98} for the Rho Ophiuchus dark cloud and 
for NGC7023, and the second to the data of \citet{Laurent00} for the M82 starburst. The second fit used more 
components, as listed in \citet{Verstraete01}, and also included a 2 component fit to the 15-20\mum\ 
C-C-C bending modes \citep{VanKerckhoven00}.  This fit applies to the Orion Bar region, M17-SW, and NGC2023. 
The final model PAH emission spectrum adopted for the MAPPINGS IIIq code is given in Table \ref{PAHtable}.

\begin{deluxetable}{lllll} 
%\tablewidth{2.0in}
\tablecaption{ Fit to PAH Emission Components\label{PAHtable}}
\tablehead{Parameters:  
}

\startdata
$\lambda~(\mu$m) & 3.39 & 6.2~(\#1) & 6.2~(\#2) & 6.2~(\#3)\\
$x_0$ &  3040.3 & 1608.5 & 1608.5 & 1593.9 \\
Peak & 1.049E-4 & 4.738E-4 & 6.768E-4 & 3.113E-4 \\
  $\sigma$ & 20.40 & 31.62 & 12.1 & 34.90 \\
  &  &  &  &  \\
$\lambda~(\mu$m) & 6.949 & 7.5 & 7.6 & 7.8 \\
$x_0$ &  1440 & 1328.5 & 1311.3 & 1274.3 \\
Peak & 3.214E-4 & 6.768E-4 & 9.730E-4 & 1.438E-3 \\
  $\sigma$ & 100 & 44.72 & 14.0 & 35.0 \\
&  &  &  &  \\
$\lambda~(\mu$m) & 8.6 & 10.0 & 11.3~(\#1) & 11.3~(\#2)\\
$x_0$ &  1163.1 & 998 & 889.4 & 889.8 \\
Peak & 1.015E-3 & 2.115E-4 & 1.015E-3 & 2.200E-3 \\
  $\sigma$ & 30.0 & 158 & 8.8 & 5.68 \\
  &  &  &  &  \\
$\lambda~(\mu$m) & 11.3~(\#3) & 12.7& 13.6 & 14.3 \\
$x_0$ &  880.9 & 787 & 735 & 699 \\
Peak & 1.354E-3 & 1.015E-3 & 1.184E-4 & 1.099E-4 \\
  $\sigma$ & 15 & 30.1 & 14.9 & 14.9 \\
  &  &  &  &  \\
$\lambda~(\mu$m) & 16.4 & 17.8 & &  \\
$x_0$ &  610 & 562 &  &  \\
Peak & 4.230E-4 & 2.961E-4 &  &  \\
  $\sigma$ & 17 & 30.0 & &  \\
\enddata

\end{deluxetable}

Emission within the MAPPINGS IIIq code is treated as an energy conservation process. In equilibrium all 
the energy gained by a PAH through the absorption of photons can be either lost through photoelectric
processes or IR emission. The fraction of the energy lost through photoelectric processes is determined,
and we assume the remaining fraction of energy is re-emitted in the IR according to our empirical fit to the 
astrophysical PAH emission bands (UIBs). 

This is not an exact treatment as the PAH molecules will likely undergo stochastic heating processes and will 
lose some of their energy through IR continuum emission instead of \emph{via} these fluorescent bands. However 
the code also allows for very small graphite grains which do undergo stochastic heating, thus both forms of 
emission are catered for.

\subsection{PAH Photodestruction}
Where detailed measurements have been made of the distribution of PAH emission within compact \HII\ regions,
it is found that the PAH avoids the ionized regions, and is instead found in a narrow zone of emission
(the photo-dissociation region) beyond the outer boundary of the ionized gas \citep{Burton00}. This
is clear evidence that PAH molecules cannot survive for significant lengths of time in the hostile 
environment in the ionized zones of an \HII\ region. This is to be expected, because the 
photodissociation timescales in this region are very short. However, PAHs can survive, and are excited into
emission in the photodissociation regions (PDRs) outside the ionized \HII\ region.

The condition for photodestruction may be set in a number of ways. The simplest is that the presence of 
hard photons leads to rapid photodissociation in the ionized gas. In this case, the PAHs are destroyed
in the ionized gas and their C is returned to the gas phase.

A somewhat more sophisticated approach would be to say that PAHs are destroyed whenever the absorption
weighted mean radiation field (not necessarily ionizing) results in a photodissociation timescale which 
is short compared to the residence time of the PAHs in that radiation field. In this case we would compare
photodissociation timescales with dynamical timescales to obtain a criterion for survival. 

A still more sophisticated approach posits the possibility that photodissociation (by the ejection of an 
acetyleneic group) is countered by repair through accretion of carbon atoms \citep{Allain96a,Allain96b}. 
In this case, if $\tau _{\rm diss}$ is the radiative dissociation timescale, and $\tau _{\rm acc}$ is the 
C atom accretion timescale then 

\begin{equation}
\tau _{\rm diss}=\left( F_{{\rm FUV}}\sigma _{\rm diss}\right) ^{-1},
\label{eqn15}
\end{equation}
and 
\begin{equation}
\tau _{\rm acc}=\left( n_{{\rm H}}X_{{\rm C}}k_{\rm acc}\right) ^{-1},
\label{eqn16}
\end{equation}
where $F_{{\rm FUV}}$ is the far-UV radiation field, $\sigma _{{\rm diss}}$ is the effective photodissociation 
cross section per PAH molecule for this particular radiation field spectrum, $n_{{\rm H}}$ is the number 
density of hydrogen atoms, $X_{{\rm C}}$ is the abundance of C in the ISM, and $k_{{\rm acc}}$ is the reaction 
rate for sticking of a carbon atom onto such a molecule. For the far-UV radiation field, we take the total
radiation field above the ionization potential adopted for these molecules ($\ge 6$~eV). 
This provides the condition for PAH destruction \citep{Dopita02}:
\[
F_{{\rm FUV}}\sigma _{\rm diss}>n_{{\rm H}}X_{{\rm C}}k_{{\rm acc}},
\]
or 
\begin{equation}
{\cal H}=\frac{F_{\rm FUV}}{cn_{\rm H}}>\frac{X_{{\rm C}}k_{{\rm
acc}}}{\sigma _{\rm diss}}.  \label{eqn17}
\end{equation}
where ${\cal H}$ is the {\em Habing Photodissociation Parameter}, defined by analogy with the dimensionless 
Ionization Parameter ${\cal U}$ used in \HII\ region theory. From observation, values of ${\cal H} \sim 0.05$
are observed in regions which still contain PAHs \citep{Allain96a}. However, this is a line of sight average, 
which provides only an upper limit. A threshold of ${\cal H}\sim 0.005$ for the destruction of PAHs ensures 
that PAHs are excluded from the \HII\ regions, but are present in
the PDRs of even the compact \HII\ regions. However, similar results are obtained by assuming that PAHs are
destroyed on a short timescale whenever the spectrum-averaged photodissociating field exceeds about ten times
the local interstellar field. Both of these formulations are available within the MAPPINGS~IIIq code. In this
paper we have opted to use the Habing Photodissociation Parameter formulation with ${\cal H} \sim 0.005$,
which effectively excludes PAH from surviving within the ionized gas.

When, locally, the threshold of ${\cal H}\sim 0.005$ is exceeded, we have assumed that the PAHs are destroyed
in our models, and in this case we return the Carbon that they contain to the gas 
phase. In starburst galaxies, this condition is sufficient to ensure that the PAHs are destroyed in both the 
ionized and in the warm diffuse media. Since the PAH absorption cross section is like that of carbonaceous grains, 
this means that most of the carriers of the 2200\AA\ absorption are removed from lines of sight which permit
the UV stellar continuum to escape. As we show below, this provides a natural explanation of why the 2200\AA\ 
feature is weak or absent in the UV spectra of AGN or starburst galaxies \citep{Calzetti01}, provided that  
small organic grains or other carbonaceous types are not present in high abundance.

\section{Grain Size Distribution \& Composition} \label{size}

The grain size distribution and the grain composition are the principal factors which determine the 
wavelength-dependence of the absorption and scattering processes. The grain size distribution in the 
ISM results from the balance between the grain formation and destruction processes. It has usually 
by a power law over a wide range of sizes, $a$, \citep{MRN77};
\begin{equation}
dN(a)/da = k a^{-\alpha} \quad a_\mathrm{min} \le a \le a_\mathrm{max}. \label{MRN}
\end{equation}

where $\alpha$, $a_\mathrm{min}$ and $ a_\mathrm{max}$ are derived by an empirical fit to the 
scattering and extinction in the local ISM. For the MRN distribution the minimum size is set to
$a_\mathrm{min}=0.005$\mum\ for carbonaceous grains, and to $a_\mathrm{min}=0.01$\mum\ for
silicaceous grains. For both grain types, $a_\mathrm{max}=0.25$\mum.

A more complex distribution has been suggested on physical grounds \citep{Weingartner01a}, and this takes
into account both silicate and carbon-containing grains with different grain-size distributions. For the
carbonaceous grains, the PAHs are treated as providing an additional component on a continuous 
distribution of grain sizes. Even these distributions can be approximated by power-laws over a wide range 
of radii. 

Grain shattering has been shown to lead naturally to the formation of a power-law size distribution of 
grains with $\alpha \sim 3.3$ \citep{Jones96}, observationally indistinguishable from the MRN value,
$\alpha \sim 3.5$. Such a grain size distribution can also hold only between certain limits in size,
the smaller size being determined by natural destruction processes such as photodestruction, and the
upper limit being determined by limits on the growth by condensation and sticking. To capture these
elements of the physics, we have adopted a modified grain shattering profile with the form;
\begin{equation}
dN(a)/da = k  a^\alpha
\frac{e^{-(a/a_\mathrm{min})^{-3}}}{1+e^{(a/a_\mathrm{max})^3}}. \label{GSprof}
\end{equation}

This creates a smooth exponential cut-off in terms of the grain mass at both the minimum and
maximum grain size of the distribution. This smooth cut removes any edge effects in either the emission 
or extinction by dust that arise due to the sharp cut-offs in the other distributions. Our distribution 
does not have the extra bump to accommodate the PAHs, since they are treated as a separate molecular 
component, as described above

The grain size limits for both graphitic and silicaceous grains are set to $a_\mathrm{min}=0.004$\mum, 
and $a_\mathrm{max}=0.25$\mum. The somewhat smaller size cutoff than the MRN distribution provides
a stronger dust re-emission continuum below 25 \mum, which is needed to fit the observed IRAS colors of 
starbursts \citep{Rush93}. 
The constant k is determined by the total dust-to-gas mass ratio, which is determined by the depletion of 
the heavy elements onto dust. Our models use a solar abundance set (as modified by the latest abundance 
determinations \citet{Allende01,Allende02,Asplund00a,Asplund00b}). The depletion factors are those used by 
\citet{Dopita00} for starburst and active galaxy photoionization modeling  and are similar to those found by 
\citet{Jenkins87}  and  \citet{SavageS96} in the local ISM using the UV absorption lines to probe various 
local lines of sight. 

Within the MAPPINGS IIIq code, the dust grain size distribution is divided into 80 bins
spaced logarithmically between 0.001--10  \mum. The number of grains of each type in each bin
is then determined by equation \ref{GSprof}. The absorption, scattering and photoelectric heating 
is calculated for each size bin. This is then used to establish the temperature probability distribution
for computation of the FIR re-emission spectrum with quantum fluctuations. 

When present, the PAHs are set at an abundance which uses 110ppm of Carbon. Consequently, the carbonaceous 
grain component is present at only rather low abundance; equivalent to 69ppm of Carbon.

\begin{deluxetable}{lll} 
%\tablewidth{2.0in}
\tablecaption{ The Solar Abundance Set ($Z_{\odot}$) and 
 logarithmic depletion factors log(D) 
 adopted for each element.\label{Z_table}}
\tablehead{
\colhead{Element} 
& \colhead{$\log({\rm Z_{\odot}})$}
& \colhead{$\log({\rm D})$}\\
}
\startdata
H & ~0.00 & ~0.00 \\
He & -1.01 & ~0.00 \\ 
C & -3.59 & -0.52 \\
N & -4.20 & -0.22 \\
O & -3.34 & -0.22 \\
Ne & -3.91 & ~0.00 \\
Na & -5.75 & -0.60 \\
Mg & -4.42 & -0.70 \\
Al & -5.61 & -1.70 \\
Si & -4.49 & -1.00 \\
S & -4.79 & -0.22 \\
Cl & -6.40 & -0.30 \\
Ar & -5.44 & ~0.00 \\
Ca & -5.64 & -2.52 \\
Fe & -4.55 & -2.00 \\
Ni & -5.68 & -1.40 \\
\enddata
\end{deluxetable}

\subsection{Grain Quantum Fluctuations}
In order to properly model the FIR emission bump, it is very important to take stochastic quantum heating of the 
dust grains into account. In particular, the  temperature fluctuations of grains serves to provide a population 
of small grains which are hotter than their equilibrium temperature and thus an excess in their IR emission 
at shorter wavelengths. 

This idea of the temperature fluctuations in small grains can be extended to the very small grains or molecules, 
such as Polycyclic Aromatic Hydrocarbons (PAHs). These are much more likely to undergo such a fluctuation, 
although the possibility of destruction by a singe photon is also increased. The point at which a dust grain, 
which cools through a continuous radiative decay, should be considered as a dust molecule, which 'cools' 
(enters lower vibrational states) through discrete vibrational and rotational quantum states is rather ill-defined.
Here we have taken the simple approach and treated all dust as grains having well-defined bulk properties, 
with the molecular treatment being reserved for the PAH population.

Temperature fluctuations and stochastic heating processes have been considered by several authors since 
Greenberg first suggested these ideas \citep[e.g.][]{Duley73,Greenberg74,Purcell76,Dwek86}.
The most detailed work on dust and temperature fluctuations has been a series of papers by Draine \& co--authors
\citep{Draine85,Guhatha89} culminating in a set of papers with Li \citep{Draine01,Li01}.

The FIR emission treatment used in the MAPPINGS IIIq code, whose results are presented here is based on the 
algorithms of \citet{Guhatha89} (GD89) and \citet{Draine01} (DL01). The code provides an optimized
solution of their algorithms to determine the dust grain temperature distributions for each grain size 
according to the both the strength and detailed spectrum of the local radiation field. The code then
integrates the resultant FIR emission of the ensemble of dust grains to provide the local re-emission
spectrum, which is itself integrated through the model in the outward-only approximation.

\section{The Attenuation by Dust; IR to UV} \label{attenuation}
If our dust model is really representative of the dust found in starburst galaxies, then it should 
reproduce the wavelength-dependent attenuation law seen in starburst galaxies \citep{Calzetti01}. 
Here we use the word  `attenuation' to distinguish it from extinction, since it properly describes 
the wavelength-dependent escape fraction of starlight from the galaxy averaged across all 
possible sight lines rather than the wavelength-dependence of the effective extinction optical depth 
along a single sight line.

Models of the attenuation show that it can be well-represented by the effect of a foreground screen 
\citep{Meurer97,Meurer99}. In an earlier paper, \citet{Fischera03} showed that many features of the 
Calzetti attenuation law for starbursts \citep{Calzetti01} could be reproduced by turbulent screen 
having a log-normal column density distribution. The grain model used in that paper was that of
\citet{Weingartner01a}, which, being strictly applicable to the local ISM,  gave a strong 
2200\AA\ absorption feature. However this feature is not evident in the attenuation curves of either 
starburst galaxies or AGN.

Here we make the hypothesis that the major carrier of the 2200\AA\ absorption feature are the PAHs,
which are, as we have seen, effectively destroyed within the strong UV radiation field environments of 
the diffuse neutral or ionized phases of the ISM. They do, however, persist in the PDRs around the 
boundaries of dense molecular clouds where they are strongly excited to fluoresce in the `UIB' PAH features 
in the infra-red. These molecular clouds are completely opaque to the passage of optical and UV photons,
and so provide only a geometrical covering factor to the attenuation at these wavelengths.

This model offers the potential both to explain the absence of the 2200\AA\ absorption feature 
\emph{and} the great intensity of the UIBs in starburst galaxies. Does it work?

\begin{figure*}
  \caption{\label{fig3} The extinction, absorption, scattering and g-factor (thick solid) of our dust 
   model for the diffuse phase of the ISM in which the UV field is sufficienty strong for PAHs to be 
  destroyed. This component is assumed to provide the foreground screen attenuation for the starburst.
  The absolute to relative extinction is $R_V=3.38$. The dotted curves are the corresponding values for 
  the local interstellar medium of our galaxy as defined by \citet{Weingartner01a}. The generally lower 
  opacity of our grain model mostly reflects the lower abundances associated with the revised solar 
  abundance set.}
\end{figure*}

In Figure \ref{fig3}, we show the wavelength dependence of the optical properties of our dust model
compared with that of \citet{Weingartner01a}. It is clearly apparent that the 2200\AA\ absorption feature
is strongly suppressed by our grain model, but otherwise that the extinction and scattering properties of 
our grain mix is quite similar to the \citet{Weingartner01a} results once the offset due to the abundance 
set is accounted for.

For comparison, we show the optical data for our grain model for the PDRs. Here the 2200\AA\ feature is 
very strong, and the IR opacity low. A linear combination of the model curves in Figure \ref{fig3} and 
\ref{fig4a} would produce a outcome which is almost indistinguishable from those of \citet{Weingartner01a}.

\begin{figure*}
  \caption{\label{fig4a} The extinction, absorption, scattering and g-factor for our dust model where
  110ppm of the carbon is assumed to be condensed in PAH. These curves correspond to our assumed 
		grain and PAH population in the PDRs and molecular clouds. Note the extreme strength of the 2200\AA\ 
  absorption, and the deficit in the absorption and extinction at IR wavelengths, relative to the 
  \citet{Weingartner01a} grain model (dashed curves). For this grain model the absolute to relative 
  extinction is $R_V=1.92$.}
\end{figure*}

We have used the optical parameters of Figure \ref{fig3} to construct an extinction curve for this grain model. 
This is shown as the thin solid line  on Figure \ref{fig5}. We then used the same procedure as described
in  \citet{Fischera03} to derive an attenuation curve appropriate to a turbulent foreground screen having 
a log-normal distribution in column densities and having an $R_V$-value $=4.0$ consistent with that obtained 
for the Calzetti curve ($R_V=4.04$). Our theoretical attenuation curve is shown on  Figure \ref{fig5} as the
thick solid line, for comparison with the Calzetti empirical attenuation curve for starbursts (dashed line).

It is clear that our grain model is extremely successful in reproducing the Calzetti empirical law to an
accuracy which lies within the observational errors in the empirical fit. This success relies upon both the 
destruction of the PAHs in the diffuse phases of the ISM, and upon the low abundance of other carbonaceous
grain types which was forced upon us by the need to satisfy C abundance constraints while at the same time
ensuring that the strength of the PAH emission bands matched observational constraints.

\begin{figure}
  \caption{\label{fig5}
  Attenuation curve for a turbulent dusty foreground screen (thick solid line) derived
  using our dust model. The $R_{\rm V}$-value is chosen to be $4.0$ consistent with the value 
		of the Calzetti curve ($R_{\rm V}=4.04$). The Calzetti attenuation curve is shown as dashed 
  line and is identified by name. 
		The initial extinction curve for our dust model is shown as thin solid line.
  }
\end{figure}

For the convenience of observers, we have given this theoretical attenuation law in tabular form in 
Table \ref{table3}. Data values are given in terms of the variable
\begin{equation}
  \xi_\lambda = \frac{A_{\lambda}-A_{\rm V}}{A_{\rm B}-A_{\rm V}} 
= \frac{E(\lambda-{\rm V})}{E{\rm(B-V)}}
\end{equation}
and the absolute attenuation can be calculated from these values by:
\begin{equation}
  A_{\lambda} = \left(\xi_\lambda / R_{\rm V}+1\right)\,A_{\rm V} \label{eqn21}
\end{equation}
with $R_{\rm V}=4.0$.

\begin{table}
  \caption{\label{table3} 
			Relative Attenuation $E(\lambda-V)/E(B-V)$ given for $R_V=4.0$ ($A_{V}=1$).}
  \begin{center}
  \begin{tabular}{l|c|ccc}
  \hline
  \hline
   $\lambda[\mu{\rm m}]$ & Waveband & This Work & Calzetti & Weingartner \&\\ 
 & or Feature &  & (2001) & Draine (2001)\\     
   \hline
   0.0912 & ${\rm Ly}\infty$   & 9.164 &11.033 & 8.591\\
   0.1053 &                          & 8.542 & 9.198 & 7.308 \\
   0.1216 & ${\rm Ly}\alpha$  & 7.397 & 7.765 & 5.885 \\
   0.1550 & \ion{C}{4}            & 5.999 & 6.062 & 4.310 \\
   0.1906 & \ion{C}{3}]            & 4.954 & 5.037 & 4.471 \\
   0.2175 & 2175                  & 4.839 & 4.415 & 5.619 \\
   0.2480 &                          & 3.904 & 3.790 & 4.025 \\
   0.3000 &                          & 2.759 & 2.862 & 2.729 \\
   0.3650 & U                       & 1.839 & 1.896 & 1.847 \\
   0.440   & B 			    & 1.000 & 1.000  & 1.000\\
   0.458   & V 			    & 0.000	& 0.000 & 0.000\\
   0.720   & R 			    & -1.177 & -1.111 & -1.103\\
   1.030   & I 			    & -2.432 & -2.267 & -2.211 \\
   1.239   & J 			    & -2.905 & -2.720 & -2.626\\
   1.649   & H 			    & -3.401 & -3.275 & -3.105 \\
   2.192   & K 			    & -3.682 & -3.690 & -3.444\\
     \hline
   \hline
  \end{tabular}
   \end{center}
\end{table}

\section{Modelling the Starburst}\label{Staburst}
We assume that the starbust region occupies a roughly spherical region filled with \HII\ regions of all possible
ages - the exact shape of the starburst region does not alter the results of the modelling. At any age, 
these \HII\ regions have a radius determined by both their age and the pressure in the ISM, according to 
Figures \ref{fig1} and \ref{fig2}. The region around the starburst is assumed to be filled with a warm, dusty 
PAH-free diffuse ISM which provides a foreground screen extinction as described in the previous section.
The starburst is assumed to form stars at a rate of $1$~M$_{\odot}$~yr$^{-1}$ for a total period of 
$10^8$~years, and since each cluster has $10^4$~M$_{\odot}$ of young stars, there are 1000 \HII\ regions
younger than 10 Myr. After 10 Myr, the ionizing photon output of the central cluster has dropped to negligible 
levels. Stars older than 10Myr are distributed within the \HII\ regions in proportion to the volumes of the
individual \HII\ regions. The \HII\ regions are assumed not to be `leaky' to the EUV photons. That is to say,
they fill their Str\"omgren spheres and so absorb all the ionizing photons. 

The input stellar spectra (stellar cluster + old stars) within the volume of an \HII\ region of given age
are computed using the STARBURST99 code as described above. The MAPPINGS~IIIq code is used with these 
stellar spectra to compute the radiative transfer through the ionized gas, the equilibrium ionization and 
temperature, and the dust and nebular line and continuum spectra throughout the individual \HII\ regions. 
The MAPPINGS~IIIq models have an isobaric density structure, where the internal pressure is defined by 
our evolutionary modelling (see figure \ref{fig2}), and their physical size matches the computed diameter 
(see figure \ref{fig1}).  The integrated output spectra are weighted by the number of \HII\ regions in each 
time step, to give the flux equivalent to a galaxian mean star-formation rate of $1$~M$_{\odot}$~yr$^{-1}$. 
The time sampling of the individual models is first at 0.4Myr, then at 1Myr and then at 1Myr age intervals 
up to 10Myr. 

We should note here that the MAPPINGS~IIIq code treats the radiative transfer in the 
outward-only approximation, scattered photons being divided into an outward stream and a back-scattered 
stream which is assumed to return across the ionized nebula.

Following \citet{Silva98} and \citet{Popescu00}, we consider each \HII\ region to be surrounded by an 
incomplete opaque shell of molecular gas, in pressure equilibrium with the ionized plasma, which blocks 
a fraction of the incident stellar light according to the geometrical covering factor. The PDR at the inner 
boundary of the molecular shell absorbs which almost all the incident FUV, UV and optical photons. 
To ensure that this is the case, the models were continued into the PDR until an \HI\ column of  
log$(n_{\rm HI}) = 21.5$ was reached. 
This corresponds to $A_V \sim 3$mag -- equivalent to nearly 10mag at the Lyman Limit. Thus, effectively all 
the stellar photons and photons of nebular origin which can excite the PAHs, or produce FIR continuum 
shortward of 100\mum\ are fully absorbed in the PDR. 

If the incomplete shell covers a fraction $f=\Omega/4\pi$ of the total solid angle, then in the 
remaining $1-f$ of the solid angle, both the remaining (unabsorbed) stellar photons, and the photons 
produced in the \HII\ region are assumed to escape from the starburst region. Some of these will
be lost in the outer foreground turbulent dusty screen, but the additional attenuation produced here
can be calculated using the results of Section \ref{attenuation}. The factor $1-f$ is therefore the escape 
probability for the stellar photons which successfully exit the starburst region. In this sense, 
the molecular clouds serve to provide a `grey' attenuation factor of $1-f$.

Since the hot stars are born in molecular clouds, and these clouds are dissipated during the lifetime of 
the \HII\ region as the region evolves from compact to classical shell morphology, it is clear that the 
molecular cloud covering factor $f$ is a decreasing function of time. This concept of age-selective
obscuration was first introduced by \citet{Silva98}, and employed by \citet{Tuffs04} in their calculation 
of the attenuation of starlight in normal galaxies. Here, we have investigated two cases. 
In the first, the covering factor decreases linearly with time from 0.9 at 1~Myr to 0.1 at 9~Myr, is complete
at zero Myr and equal to 0.1 at times later than 9~Myr. Since this function is a little arbitrary, we also 
investigated the more general case in which the covering factor decreases exponentially with time:
\begin{equation}
f(t)=exp\left[ -t/ \tau_{\rm clear} \right] \label{eqn22}
\end{equation}
with the molecular cloud clearing timescale, $\tau_{\rm clear}$, set to 1, 2, 4, 8, 16 and 32 Myr. It turns
out that the linear case provides an SED which is indistinguishable from the exponential case and has an 
effective timescale $\tau \sim 7$Myr. 

Since, in general, there will be no physically \emph{a priori} way of constraining the molecular cloud 
clearing timescale, we are forced to leave it as a free parameter, which can be hopefully constrained 
by the observed SED or colors of real starburst regions.

\begin{figure*}
 \caption{\label{fig6}
  The cumulative SED for the \HII\ regions and their stars of different ages 0.4 (lower curves)-- 10 Myr 
  (top). The older stars are included within the volume of the \HII\ regions. Each contributing spectrum 
  is weighted in proportion to its covering factor $1-f(t)$ The case of the covering factor decreasing 
  linearly with time and the ISM pressure $P/k=10^4$~cm$^{-3}$K is shown here (see text). }
\end{figure*}

To show the way in which the final SEDs are composed of the sums of many \HII\ region models, and to 
demonstrate that all models provide important contributions to the final SED, we display the cumulative 
SEDs resulting from this addition for \HII\ regions of ages 0.4 -- 10 Myr in figures \ref{fig6} and 
\ref{fig7}. These figures apply to the case where the molecular cloud covering factor decreases linearly 
with time, and for an ISM pressure $P/k=10^4$cm$^{-3}$K. The y-axis is chosen to display the energy 
flux $\nu F_\nu$, rather than $F_\nu$, in order to better show how the FIR bump re-distributes the UV 
stellar continuum. The absolute scale for the summed SED corresponds to the star formation rate
of 1.0~M$_{\odot}$~yr$^{-1}$ used in the models.

In figure \ref{fig6}, which shows the fraction of flux escaping from the \HII\ regions, note the strength 
of the nebular lines and continuum, the weakness or absence of PAH features, the hot dust temperatures and 
the presence of silicate absorption in the infrared spectra of the youngest \HII\ region models. For 
these models, the stellar continuum dominates only in the UV. In the visible and near-IR the nebular 
continuum dominates, with the bound-free edges prominent. 

After about 3Myr, the red supergiant stars appear with their characteristic CO features in the H and K 
bands. For these models of older clusters, the stellar continuum dominates the SED up to about 5\mum\ .

In figure \ref{fig7}, the additional contribution to the SED by the \HII\ region in the remaining solid angle
and by the photodissociation regions is shown. In these spectra, all the stellar the UV continuum radiation 
has been absorbed in the PDR and converted to dust re-emission, which thanks to the low optical depth in the 
IR is free to escape. Note how the peak of the FIR dust re-emission bump shifts to longer wavelengths for 
the models with the older central clusters. This is a consequence of both the decrease in the stellar 
luminosity and the larger shell sizes, which both lead to lower dust temperatures. Also, the PAH emission 
bands are stronger in the older \HII\ regions. In these models, the PAHs survive thanks to the lower 
photo-destruction rates of the PAHs resulting from the lower UV fluxes and cooler stellar temperatures. 
Nonetheless, there are still plenty of photons with $6 \le h\nu \le 13.6$eV to be absorbed by the PAHs and 
to excite the UIB features.

Thus our models can succeed in explaining the apparently contradictory observations that starburst galaxies 
are seen to have very strong UIB emission (implying the presence of large quantities of photon-excited 
PAH molecules), but have no sign of the UV 2200\AA\ absorption feature (implying the absence of PAHs, 
or at least carbonaceous grains, in the more diffuse phases of the ISM).

\begin{figure*}
 \caption{\label{fig7}
  The cumulative SED of figure \ref{fig6} (lower curve) to which has been added the contributions
  of the optically thick portions of the \HII\ regions and their PDRS with ages 
	 from 0.4 (lower curves) to 10 Myr (top). As in figure \ref{fig6}, the case of the ISM pressure 
  $P/k=10^4$cm$^{-3}$K is shown, and the case for which the covering factor changes linearly with time 
		(see text).	Each contributing spectrum is added with a weight in proportion to its covering factor $f(t)$.}
\end{figure*}

\section{Parameters Controlling the SED} \label{SED}
In this section, we present the results of the systematic investigation of the two parameters which, apart
from the optical depth of the foreground absorbing screen, the assumed dust grain size distribution
and the assumed dust composition, primarily control the SED of the starburst. These parameters are identified
as the mean pressure in the ISM, and the molecular cloud covering factor.

\subsection{The Role of Pressure} \label{pressure}
We have computed three sets of models with different pressures, $P/k=10^4, 10^6$ and $10^7$cm$^{-3}$K. 
The first of these corresponds roughly to the pressure in the ISM near our sun \citep{Jenkins83}, and so 
would be applicable to star formation complexes in disk galaxies.

Pressures of $P/k\sim 10^6$ are commonly encountered in starburst galaxies, and can be measured using either the 
emission measure in the \HII\ regions or even using the density-sensitive 
[\ion{S}{2}] $\lambda \lambda 6717/6731$\AA~ ratio, provided that $n_e \ge 100$~cm$^{-3}$. Starbursts with
these pressures are characterized by warmer IRAS colors \citep{Kewley01,Kewley01b}.

Finally, starbursts with $P/k \sim 10^7$ have the warmest IRAS colors, and are most often the nuclear starbursts
in post-merger galaxies. For these, the [\ion{S}{2}] $\lambda \lambda 6717/6731$\AA~ ratio indicates 
$n_e \ge 1000$~cm$^{-3}$ \citep{Kewley01,Kewley01b}. Thus, our choice of parameters should cover the full 
range seen in real starburst galaxies.

Figure \ref{fig8} shows the effect of the changing pressure on the SEDs. The main change is the progressive 
shift of the peak of the FIR dust re-emission bump to shorter wavelengths, and the increase in its flux.
For a pressure of  $P/k=10^4$, the peak is at 105\mum, at $P/k=10^5$, the peak is at 70\mum\ and by
$P/k=10^7$, the peak has shifted down to 50\mum. At this high pressure, hot dust emission can be traced
to as wavelengths as short as 4\mum.

\begin{figure*}
 \caption{\label{fig8}
  The summed starburst SEDs computed for ISM pressures  $P/k=10^4, 10^6$ and $10^7$cm$^{-3}$K. 
   Computed for the case in which the molecular cloud covering factor changes linearly with time.}
\end{figure*}

For all these curves, there is no discernable change in the stellar continuum. This is to be expected as it 
is defined by the geometrical constraints. However, there is a progressive change in the nebular line 
spectrum towards higher excitation at higher pressure. This is a consequence of the mean ionization 
parameter, which slowly increases as the pressure increases.

Surprisingly enough, the strength of the PAH emission features is almost invariant with pressure. This 
arises because the PAH molecules can survive in the PDRs independently of the pressure, since the 
photodissociation rates are low in these zones. Thus, the PAH features scale only with the covering
factor, $f$.

We would expect these theoretical SEDs to be not very accurate at wavelengths $>100 \mu$m because we 
have not taken into account either the dust re-emission of diffuse starlight, or caused by stars older 
than $10^8$years. Nor have we included the FIR re-emission by foreground dust outside the main 
star-forming region. Thus, we would expect that in real starburst galaxies there will be a cold component 
of dust emission from the ambient medium which is not predicted by this model. However, for starburst 
galaxies the results in Section \ref{Obs} shows the fit to be remarkably good. Nonetheless, the absence of 
this component will certainly become a serious problem for the modelling of ``normal" (non-starburst) 
galaxies, since we know that for these galaxies the luminosity of the cold dust component generally 
exceeds that of the locally heated dust \emph{e.g}. \citet{Popescu02}. High resolution maps of galaxies 
like M31 \citep{Haas98}, M33 \citep{Hippelein03} or NGC891 \citep{Popescu04} made in the ISO 170 or 200\mum\
bands all show that most of this cold dust luminosity is carried by the diffuse ISM, not by the optically 
thick molecular clouds. Thus, the model presented in this paper is not really applicable to normal galaxies. 
For this, the output SED of the HII regions computed here will have to be incorporated in a model such as
that of \citet{Popescu00} that self-consistently calculates the radiation transfer in the diffuse 
component of the ISM.

Another way in which these theoretical SEDs may differ from those of real starburst galaxies is in the 
the way collective effects may alter the form of the computed SED. In galaxies with a strong centrally
concentrated starburst, \HII\ regions may be superposed along the line of sight, and the radiative transfer
would be then more complex than assumed in our model. Again, the solution to this is to build a 
three-dimensional radiative transfer model for the dusty gas, and to use the SEDs presented here as input
source functions.

Finally, there may well be a population of dust-enshrouded and confined ultra-compact \HII\ regions 
contributing to hot dust emission. We might reasonably expect to find relatively more of these in 
high-pressure regions. In our own Galaxy the W3 complex provides precisely these conditions. At high pressure, 
the star-forming regions are embedded in denser molecular gas dust, and the \HII\ regions stall sooner, 
and at smaller radii. Single OB stars would therefore find it difficult to blow their \HII\ regions to
sufficient size to be able to communicate with regions of lower density, and may therefore spend a 
significant part of their main-sequence lifetime in the Compact or Ultra-Compact \HII\ region phases. 
We will return to the evidence for such a population of compact \HII\ regions in Section \ref{Obs}, below.

\subsection{The Role of Covering Factor} \label{covering}
The molecular cloud covering factor has a significant effect on nearly all the starburst SED, as Figures
\ref{fig9}, \ref{fig10}, and \ref{fig11} make clear. These figures show the effect of the exponential 
molecular gas clearing timescale, $\tau_{\rm clear}$ at the three different pressures.

\begin{figure*}
 \caption{\label{fig9}
  The summed starburst SEDs computed for ISM pressure  $P/k=10^4$cm$^{-3}$K with exponential
  molecular cloud clearing timescales of 1, 2, 4, 8, 16 and 32 Myr. Note that a long clearing 
	 timescale depresses the UV continuum but leads to stronger PAH emission as well as stronger 
		$\sim$ 100$\mu$m dust emission.}
\end{figure*}

\begin{figure*}
 \caption{\label{fig10}
  As Figure \ref{fig9}, but with  $P/k=10^6$cm$^{-3}$K. Note the displacement of the FIR bump
towards shorter wavelengths.}
\end{figure*}

\begin{figure*}
 \caption{\label{fig11}
  As Figure \ref{fig9}, but with  $P/k=10^7$cm$^{-3}$K.
		}
\end{figure*}

First, high covering factors lead to an attenuation of all the stellar continuum, which is dominant 
from the Lyman Limit down to about 5\mum. However, when the molecular gas curtain is slow to open, 
there is a significant effect of ageing in the stellar continuum, since the light from the oldest 
\HII\ regions is dominant, and these \HII\ regions, being largest, also contain the largest number 
of stars with ages $> 10$Myr. This ageing is visible as an increase in the size of the Balmer and 
Paschen discontinuities, and a flattening of the UV spectral slope.

Larger covering factors also lead to an increase in strength of the PAH features, and both a strengthening
and broadening of the $\sim 100 \mu$m dust emission peak. This strengthening of the peak works in the 
opposite direction from pressure, since the strengthening is due to an increasing contribution of the older
\HII\ regions with longer $\tau_{\rm clear}$. These older \HII\ regions are larger, and consequently the 
dust in them is cooler.

At about 25\mum, the flux changes little with $\tau_{\rm clear}$ for all pressures. However, the flux at
this wavelength is very strongly affected by affected by changes in pressures. By contrast, the stellar 
continuum and the PAH features are strongly affected by the molecular cloud covering factor, but changes
in pressure do not affect these at all. This offers the potential to distinguish observationally between 
these two parameters, as we will discuss below.

\subsection{Approximations to the Nonthermal Emission} \label{radio}
One of the results of our modelling is that  the 60\mum\ flux depends quite strongly on the pressure in the 
interstellar medium. However, we also know from observation that there is an extraordinarily close correlation
between the 60\mum\ infrared continuum and the radio 1.4~GHz continuum of star forming galaxies. This linear
correlation spans  $\sim 5$  decades of magnitude with less than 0.3~dex dispersion \citep{Yun01,
Wunderlich87}, and the IR -radio correlation persists even when measurements of the cold dust emission 
(which carries the bulk of the FIR luminosity for normal galaxies) are included in the total FIR luminosity
\citep{Pierini03}. The mean relationship between the 60\mum\ flux and the 1.4~GHz continuum is:
\begin{equation}
\left[{L_{\rm 1.4GHz}}\over{\rm W~Hz^{-1}}\right]=10^{12}\left[{L_{\rm 60\mu m}}\over{L_\odot}\right]
\label{radIR}
\end{equation}

We know that, in star forming galaxies at 1.4~GHz, the non-thermal emission by relativistic electrons dominates 
by at least an order of magnitude over the free-free emission \citep{Condon92,Niklas97,Dopita02}. Therefore, 
and somewhat remarkably, this relationship couples a purely thermal process with a non-thermal process, over 
many decades of flux, and locally, within individual galaxies. Because of the intrinsic interest of this radio : FIR
correlation, and because we have shown that ISM pressure directly affects one of the parameters that enters into 
this correlation, in this section we will investigate a `toy' model of the synchrotron emission in order to gain 
insight as to how the synchrotron emission might be similarly affected by the ISM pressure, thus keeping the 
radio : FIR correlation as tight as is observed.

If the lifetime of the synchrotron electrons is short compared with the evolution timescale of the starburst,
then as \citet{Bressan02} have shown, the synchrotron emissivity acts as a bolometer of the supernova rate,
since the relativistic luminosity is ultimately derived from the Fermi acceleration process in 
supernova shocks. Specifically, the non-thermal luminosity is given by 
\begin{equation}
L_{\rm NT}(\nu) \propto \dot{M}_{*} B^{\alpha-1}\nu^{-\alpha} \label{NT1}
\end{equation}
where $\dot{M}_*$ is the total star formation rate.
Since the observed spectral index lies in the range, $0.5 > \alpha > 1.0$, there is only a very weak 
dependence on the magnetic field in this case.

If the lifetime of the relativistic electrons is long compared with the length of the starburst, then the 
emissivity of the relativistic electrons depends on the square of the local magnetic field, and the local 
density of the relativistic electrons. \citet{Groves03} ascribes the overall radio-FIR correlation, and its 
local variation in individual galaxies to the operation of magneto-hydrodynamic turbulence.

In a magnetized and turbulent interstellar medium, numerical simulations (\emph{e.g.}~\citet{ChoVish00b}) 
have shown that the local magnetic field  will stay in equipartition with the gas turbulent energy;
\begin{equation}\label{equipartition}
\delta V \sim B/\sqrt{4\pi\overline\rho}.
\end{equation}

This provides the coupling between the gas density and  the magnetic field  which is a necessary,  
but not a sufficient, condition for the operation of the radio -- FIR correlation. This equation
also ensures that the magnetic field pressure and the gas pressure are correlated.

The local synchrotron emissivity at frequency $\nu $ is given from the standard synchrotron theory by
\begin{equation} 
j_{\nu }=f(a)kB^{\left( a+1\right) /2}\nu ^{-\left(a-1\right) /2}  \label{NT2}
\end{equation} 
where $f(a)$ is a complex numerical factor of $a$, the power-law slope of the non-thermal electron 
the number density of the relativistic electrons with relativistic $\gamma $; $N(\gamma)^{-a}=k\gamma ^{-a}$, 
and $B$ is the local magnetic field. The total density of the relativistic electrons is therefore 
$n_{\mathrm{cr}}=k\gamma  _{\min  }^{1-a}/\left(  a-1\right)$, where $2< a < 3$  corresponds  
to  a frequency  spectral index of synchrotron emission $0.5 > \alpha > 1.0$. Provided that the
relativistic electrons cannot escape from the star-forming region, the constant, $k$, should be  
proportional to the star formation rate. Equation \ref{NT2} predicts, by contrast with equation \ref{NT1},
there should be a strong correlation between the emissivity and the magnetic field, in the range
$B^{3/2}$ to $B^2$.

At normal ISM pressures, the lifetime of the relativistic electrons should be longer than the 
star formation timescale, so that equation \ref{NT2} would apply. However, as the pressure increases, 
the emissivity of the non-thermal emission also increases, and the lifetime of the electrons becomes shorter.
Thus, at high $P/k$, we expect that equation \ref{NT2} ceases to apply, and we go over to equation \ref{NT1} 
instead. Unfortunately we have no way of estimating at which point this occurs with this simple theory.

In our model for the emissivity, we  have assumed that equation \ref{NT2} applies, and have adopted the case 
$a=2$, which provides a spectral index of $\alpha=0.5$ and a pressure dependence of the emissivity arising 
from the $B-$field of $P^{3/4}$. The calibration of the non-thermal flux was accomplished empirically 
by applying equation \ref{radIR} at a pressure of $P/k=10^4$cm$^{-3}$K. The emissivity was then computed for 
the higher pressures, still assuming that equation \ref{NT2} applies.

\begin{figure*}
 \caption{\label{fig12}
  The summed SEDs ($F_{\nu}$ \emph{vs.} $\lambda$) computed for ISM pressures  $P/k=10^4, 10^6$ and 
  $10^7$cm$^{-3}$K for the case in which the molecular cloud covering factor changes linearly with time
		and for an absolute star formation rate of $1.0M_{\odot}$yr$^{-1}$. Note that changes in the radio 
  non-thermal luminosity with pressure are positively correlated with changes in the 60\mum\ flux. }
\end{figure*}

The result is shown in figure \ref{fig12}. Here we have plotted the computed $F_\nu$ from the Lyman Limit 
up to 30cm (0.98GHz). The apparent change in spectral index of the radio continuum with pressure is due to
the underlying contribution of the radio free-free continuum from the ionized plasma in the \HII\ regions. 
This is relatively more important at low ISM pressure. Changes in the radio non-thermal luminosity with 
pressure are correlated with changes in the 60\mum\ flux. However, it is clear that the size of the increase 
in 1.4GHz continuum computed using equation \ref{NT2} is appreciably larger than the corresponding increase 
in the 60\mum\ flux.It seems clear that when the pressure of the ISM reaches $P/k \sim 10^6$cm$^{-3}$K, the 
lifetime of the synchrotron electrons is becoming, or has become, shorter than the age of the starburst.

Once this condition is reached, then the synchrotron flux becomes more sensitive to the instantaneous 
injection rate of relativistic electrons. We would therefore expect more dispersion in the radio -- farIR
correlation at the highest pressures. Since many of the ULIRG galaxies have high pressure, this could
contribute to scatter at the upper end of the radio -- farIR correlation \emph{c.f.}~\citet{Bell03}.

\section{Comparison With Observations}\label{Obs}

\subsection{Pan-spectral fits to Starburst Galaxies}\label{SEDfits}

In this section we compare the theoretical SEDs with the observations of two exceedingly well-observed
objects; Arp~220 ($z=0.0182$) and NGC~6240 ($z=0.0245$). These objects are chosen because the SEDs of these 
objects have been established from the Lyman Limit up to beyond 1000\mum.  Arp~220 is widely used as the 
template for violent and highly extinguished starbursts and for the computation of the fluxes expected from 
high redshift objects. The case of NGC~6240 (IRAS~16504+0228) is not so clear-cut. It is is a very well-known 
ultra-luminous infrared merger remnant with a strong non-thermal radio excess. It is clearly dominated by a 
starburst, and shows a very strong Balmer lines but has a LINER or Sy2 nuclear spectrum 
\citep{Kewley01b,Goldader97,Kim95,Veilleux95}. It also shows evidence for an obscured AGN at X-ray and radio 
wavelengths \citep{Ikebe00,Risaliti00, Kewley00}.

For both galaxies the UV and optical fluxes were taken from the third reference catalogue of bright 
galaxies v3.9 \citep{deVaucouleurs91cat}.  The near-IR ($JHK$-band) points are from \citet{Spinoglio95apj}, 
and include aperture corrections to allow a direct comparison with the larger aperture mid- and far-IR fluxes. 
The $L$-band magnitude for NGC~6240 \citep{Allen76apj} has been similarly corrected by scaling with the same 
factor as required for the $K$-band. The 3-1500\,\micron\ data points were taken from 
\citet{Klaas97aa,Klaas01aa}, \citet{Benford99thesis}, \citet{Spoon04aa} and references therein.  
All UV to near-IR fluxes have been corrected for Galactic extinction using the E(B-V) values based upon 
IRAS 100~\mum\ cirrus emission maps \citep{Schlegel98apj} and extrapolating following \citet{Cardelli89apj}.

The process of matching the models to the observations required two steps. First, the data were shifted to 
rest-frame frequencies and the absolute flux determined from the luminosity distance of the galaxy was scaled 
to the far-IR section of the SED.  The location of the peak and the slope of the long-wavelength tail of the 
thermal dust re-emission fixes the appropriate value of $P/k$. The molecular cloud clearing timescale 
$\tau_{\rm clear}$ was best determined by the fit in the 2-5 \mum\ region, which is dominated by the stellar 
continuum leaking out between the molecular clouds.  The choice of clearing timescale also changes the inferred 
scaling factor in the far-IR, but to a limited degree. The final scaling factor to be applied to the observed 
absolute fluxes is a direct measure of the star formation rate in the galaxy concerned, since all the theoretical 
spectra are scaled to an absolute star formation rate of $1.0M_{\odot}$yr$^{-1}$. To estimate this scaling factor 
we assume a flat universe with $H_{0} = 71 \kmps\,\rm Mpc^{-1}$, $\OmM = 0.27$ and $\OmL = 0.73$ 
\citep{Spergel03apj,Tonry03apj}.

Second, when a best-fit match was obtained throughout the far-IR regime, the model 
SEDs were reddened according to the foreground turbulent screen attenuation law given in table \ref{table3}. 
The total attenuation was adjusted to match the integrated fluxes at optical and near-IR wavelengths. This
provides a direct estimate of $A_{\rm V}$ to the star forming region.

The family of SEDs generated for this paper and used in this section are available as tab delimited 
text files in the electronic version of the paper. Likewise, the attenuation curves from section
\ref{attenuation} are also given in the electronic version of the journal in the same form and over the 
full range of wavelength so as to facilitate their use by others. In these files, and in the fitting used
in this section, we have not included any synchrotron contributions, since these are so uncertain at 
present. Without this contribution, the model SEDs become dominated by the free-free emission (assumed 
optically thin) at wavelengths longer than about 2~mm.

The observed SEDs of Arp~220 and NGC~6240 with model fits overlaid are shown in figures \ref{fig13} and 
figures \ref{fig14}.  High pressure models with $P/k = 10^{7}$cm$^{-3}$K were required in both cases.  The SED of 
NGC~6240 could be accurately fitted with either a linearly decreasing covering factor or 
$\tau_{\rm clear} \sim 8$~Myr and with $A_{\rm V} =2.7$. However, for Arp~220 a good match to the far-IR could be
only be obtained for a model with $\tau_{\rm clear} = 100$~Myr. This implies that the starburst is enshrouded 
by an optically thick dust and molecular screen which absorbs almost all UV/optical radiation. This molecular
screen effectively acts as a bolometer. The fit in the near-IR regime requires an additional attenuation of 
$A_{\rm V} = 5.7$. It was not possible to fit the excess of flux in the UV. This component may arise from an outer 
stellar population which is not as deeply dust embedded as the main star formation region.  
If we use the value of $ N_{\rm H}/{A_{\rm B}-A_{\rm V}}=5.8 \times 10^{21}$~cm$^{-2}$
mag$^{-1}$ \citep{Bohlin78} along with $R_{\rm V}=4.0$, then our value of $A_{\rm V}\sim 5.7$~mag translates to
an inferred column density of $N_{\rm H} \sim 9 \times 10^{21}$~cm$^{-2}$. This can be directly compared with
the X-ray absorption to the nuclear source, $N_{\rm H} \sim 3 \times 10^{22}$~cm$^{-2}$ \citep{Clements2002}.
Since we would expect the central source to be more deeply dust-embedded than the more extended star formation
region, our estimate of $A_{\rm V} = 5.7$.

With luminosity distances of 78~Mpc and 105~Mpc for Arp~220 and NGC~6240 respectively, we infer star formation 
rates of $300$\Msunpyr and for Arp~220 and $120$\Msunpyr for NGC~6240. As one might expect for these
heavily dust-enshrouded objects, these are very close to the estimates derived from the bolometric far-IR 
luminosities, $L_{\rm IR}$.

\begin{figure*}
 \caption{\label{fig13}
  The global fit to the spectrum of Arp~220. This requires a $P/k=10^7$~cm$^{-3}$K, a very long clearing 
  timescale for the molecular clouds, and a foreground dust screen with $A_{\rm V}\sim 5.7$~mag.
   The fit implies an absolute star formation rate of $300~M_{\odot}$yr$^{-1}$. Also shown is the modified
   Black Body fit to the $60-500$~\mum\ region.}
\end{figure*}

\begin{figure*}
 \caption{\label{fig14}
  The global fit to the spectrum of NGC~6240. This requires a $P/k=10^7$~cm$^{-3}$K, a molecular clearing 
  timescale $\tau_{\rm clear} \sim 8$~Myr and $A_{\rm V} =2.7$. The fit implies an absolute star formation 
  rate of $120~M_{\odot}$yr$^{-1}$. Also shown is the modified Black Body fit to the $60-500$~\mum\ region.}
\end{figure*}

The fit with the observations is, in general, rather good. For Arp~220, the main differences between the
the data and the models are the excess in UV flux in the data, already alluded to, and a difference in the
observed and predicted strength of the PAH features. The data show that the strength of the PAH features are
overestimated by a factor of about two in the models. This difference can be understood as a result of the
very high star formation rate inferred for Arp~220, which implies a strong PAH- photodissociating UV field
in the star formation region. Presumably, a greater fraction of the PAHs are destroyed in this environment.

Relative to the model, Arp~220 also shows an excess of flux in the $12-30$\mum\ band which can be ascribed
to the presence of an additional warm dust component. As discussed in Section \ref{pressure}, this is most
likely due to a population of confined compact and ultra-compact \HII\ regions around single OB stars, or
small clusters. It is highly unlikely that this is due to the presence of an AGN, since at other wavelengths,
there is no evidence to support this idea.

For NGC~6240, the main difference between model and observation is that the observations also show an excess of
flux in the $10-30$\mum\ waveband. In this object, however, an AGN is undoubtedly present.

Overall, the goodness of fit over the whole range of wavelength was somewhat surprising to us. In particular, we 
would have expected to find a poor quality of fit in the $100-1000$\mum\ region of the spectrum, since 
here the emission was thought to be dominated by the cooler dust located in the more extended dusty region 
surrounding the starburst, which is not represented in our modelling. However, this can be understood 
as a consequence of the long molecular cloud clearing timescales we inferred in the fitting procedure. 
This implies that the vast majority of the UV photons are being absorbed in the PDRs surrounding the \HII\ 
regions. This is consistent with the rather long molecular cloud clearing timescales inferred for these 
objects. Because of the high pressures these regions are very dense, and consequently, thin in comparison 
with the radius of the \HII\ regions. Thus the intensity of the local radiation field is simply a function 
of the optical depth into the PDR. Since the local radiation field determines the local temperature, or 
more strictly, the temperature distribution of the dust grains, the global dust emission spectrum is a 
superposition of a wide range of dust temperatures. The hottest grains are located close to, or within, 
the ionised region, and the coolest grains are in the tail of the PDR.

Frequently, observationalists use a single temperature modified blackbody spectrum as a fit to the 
$100-1000$\mum\ spectra of starbursts. We have therefore generated such spectra with dust emissivity 
proportional to $\lambda^{-\beta}$. These have been fitted over the region $60-500$\mum.  Both
fits yield a value $\beta = 1.5$ (similar to what is usually assumed). The effective dust 
temperatures derived are $T = 37.5$\,K for Arp~220 and $T = 41.3$\,K for NGC~6240. This is in
good agreement with what has been found for single temperature fits to bright IRAS galaxies 
\citep{Dunne00mnras}. Estimates of the Bolometric flux based on such parameterizations clearly underestimate
the true value, since these fits do not take account of the emission at the shorter wavelengths. More complex
multi-temperature fits do a better job in this regard \citep[see \emph{e.g.}][]{Klaas01aa}, but all such 
fitting is fundamentally unphysical. Especially misleading is the use of an additional parameter, the 
optical depth at 100~\mum. If this is mis-interpreted as representing a real physical parameter, it 
would imply such tremendous optical depths that the starbursts would be totally invisible at optical 
wavelengths. Such is clearly not the case.

Nonetheless, the single-temperature fits \emph{are} a useful way of characterizing the form of the long 
wavelength side of the far-IR thermal emission from dust. For example \cite{Blain04} have used a sample of 
ultra-luminous galaxies in the nearby universe as well as a sample of more distant starburst galaxies 
($1 < z <5$) to find an (admittedly loose) correlation between the inferred dust temperatures and the 
luminosity of the host galaxy. In the framework of the theory presented here, this could be interpreted 
as a correlation between the star formation rate and the pressure in the ISM of the host galaxies.

\subsection{IR Color-Color Diagrams for Starburst Galaxies}

In the following sections we compute theoretical color : color diagrams for our model starbursts as
would be seen using IRAS or the Spitzer Space Observatory. We regard the computed color:color diagrams 
involving flux densities at wavelengths longer than 100\mum\ as unreliable such as those involving the
Spitzer Space Observatory MIPS 160\mum\ filter). This is because we have neglected both the old stellar 
populations and the diffuse components of the dust, both of which are important contributors at these 
wavelengths for starbursts located in more normal galaxies \citep{Popescu00}. 

\subsubsection{IRAS}
We have used the band pass data for the IRAS 12, 25, 60 and 100\mum\ bands in conjunction with our
theoretical SEDs to construct color-color diagrams. These are shown in figure \ref{fig15} .

Some homogeneous observational data is shown for comparison. We have used \citet{Rush93} data for objects 
classified as starbursts. These are plotted as $\odot$ symbols. Typical errors in this data set
amount to 0.1 to 0.2dex in terms of errors on color-color plots.

The single square represents the mean for the \citet{Rush93} Seyfert~I colors as determined by 
\citet{Dopita98}. It is clear that our theoretical curves define an upper envelope to the observed points 
on all four of these color-color diagrams. However, our detailed spectral fitting in Section \ref{SEDfits}
shows that our models are missing a warm-dust emission component in the $12-30$ \mum\ region, which is almost 
certainly caused by a population of confined compact and ultra-compact HII regions with ages less than about 
2 Myr surrounding single OB stars or small OB clusters. This could account for the offset between the 
theoretical curves and the majority of the \citet{Rush93} starbursts. However, some of the objects classified 
as starbursts by \citet{Rush93} are in fact of mixed excitation; starburst + AGN. Indeed, the fact that 
those objects that lie particularly low on one of these color-color plots also lie low on the others makes 
this conclusion secure. From the mixing lines given in \citet{Dopita04} we might infer that these objects 
have a bolometric contribution ascribable to the AGN of between 10 and 20\% that of the starburst.

It is clear that the models reproduce the intrinsic spread in colors observed in real starburst galaxies. 
In general, warmer IRAS colors correspond to higher ISM pressures, and therefore we might expect to find a 
correspondence between the compactness of the starburst and its IRAS colors. However, none of these diagrams 
is suitable for separating the two controlling parameters of the ISM pressure and the molecular cloud 
clearing timescale. 

The scatter of the points on the $F_{100}/F_{25}$ \emph{vs} $F_{60}/F_{25}$ diagram is low. However, this
can now be seen as simply a degeneracy in the parameters -- both $P$ and $\tau_{\rm clear}$~spread points
in the same direction, which also happens to lie in the same direction as the AGN : starburst mixing line.
The same also applies to the $F_{100}/F_{12}$ \emph{vs} $F_{60}/F_{25}$ diagram. The remaining two 
color-color diagrams offer better potential to separate the contribution of any AGN to the total spectrum, 
and also the pressure can be somewhat constrained.

\begin{figure*}
 \caption{\label{fig15}
  Color-color diagrams computed for the IRAS bands. The observational points come from the \citet{Rush93} 
  starbursts. The single point shows the mean colors of Seyfert I galaxies given by \citet{Dopita98}. 
  Note that our models reproduce the intrinsic range of FIR colors seen in starburst galaxies, and that the 
  grid of models in all cases defines an upper boundary around the observed points. It is clear that, if the
  theoretical grid is correct for ``pure" starbursts, then a number of the objects classified as starbursts 
  are in fact of mixed excitation, starburst and AGN.}
\end{figure*}

\subsubsection{Spitzer Observatory}
The IRAC and MIPS instruments of the Spitzer Observatory offer the potential of observationally separating 
and measuring the both the pressure and the molecular cloud covering factor, and also of estimating the 
contribution of any AGN component. 

This can be done in the following manner. First, because the IRAC 4.5\mum\ band directly measures the 
stellar continuum, the 5.7\mum\ band sees the PAH emission and the 7.9\mum\ band is completely dominated 
by the PAH emission, the ratio of fluxes in these wavebands is very sensitive to the covering factor.

Second, the MIPS 24\mum\ band is almost invariant with covering factor, but is extremely sensitive to pressure
(see Figure \ref{fig9} through \ref{fig11}). Thus, the flux in this waveband can be used in conjunction with
either any of the IRAC bands to provide a pressure discriminant. In practice either the IRAC 4.5\mum\ or 
the 5.7\mum\ bands are best for this purpose. Alternatively, longer-wavelength MIPS bands could be used,
but these tend to suffer the same degeneracy as the IRAS bands.

Finally, the presence of an AGN continuum would tend to wash out the PAH features \citep{Laurent00}, so the
ratios of fluxes in the IRAC 7.9 \mum\ and 4.5\mum\ bands and any of the IRAC bands could be used to help
determine the AGN fraction.

Some useful Color-color plots are shown in Figure \ref{fig16}.

\begin{figure*}
 \caption{\label{fig16}
  Color-color diagrams computed for the IRAC and MIPS instruments on the Spitzer Observatory. In general 
the Spitzer wavebands are more useful in separating the effects of ISM pressure from the time-dependent
molecular cloud covering factor. Increasing pressure increases the MIPS 70 and 24 $\mu$m fluxes, but leaves 
the IRAC bands unchanged, while increasing covering factor decreases the stellar continuum (the IRAC 4.5$\mu$m
flux) but increases the strength of the PAH bands, most clearly in the IRAC 7.9$\mu$m band.}
\end{figure*}

This figure shows how either the  $F_{5.7}/F_{4.5}$ \emph{vs} $F_{24}/F_{4.5}$ or the $F_{7.9}/F_{4.5}$ 
\emph{vs} $F_{24}/F_{4.5}$ color-color diagrams provide a clean separation between $P$ and $\tau_{\rm clear}$,
while the MIPS color-color diagrams, $F_{70}/F_{160}$ \emph{vs} $F_{70}/F_{24}$ or $F_{160}/F_{24}$ 
\emph{vs} $F_{70}/F_{160}$ are essential degenerate in these parameters. However, by providing a starburst
line on these diagrams, they can be used to help estimate the contribution of an AGN, once the mean 
AGN colors have been determined.

\section{Calibration of Star Formation Rates}\label{SFRs}

\subsection{Star Formation Rates from H$\alpha$}
Neglecting the internal extinction by dust within the \HII\ regions, the H$\alpha$ flux 
can be fit in terms of the molecular cloud dissipation timescale. From our models, this gives, 
to better than 1\% for $\tau \ge 2$~Myr:
\begin{equation}
log \left[ L_{{\rm H\alpha }}\over {\rm erg.s}^{-1}\right] = 41.63 +log\left[SFR \over
\rm {M_{\odot}~yr^{-1}}\right] - log[1.6 + \left(\tau \over {\rm Myr} \right)] \label{27}
\end{equation}
and for $\tau = 1$~Myr:
\begin{equation}
log \left[ L_{{\rm H\alpha }}\over {\rm erg.s}^{-1}\right] = 41.195+log\left[SFR \over 
\rm {M_{\odot}~yr^{-1}}\right]. \label{28}
\end{equation}
For the case $\tau = 1$~Myr this gives the following calibration for the star formation rate,
\begin{equation}
\left[SFR_{{\rm H\alpha }} \over \rm {M_{\odot}~yr^{-1}}\right]
=6.4\times 10^{-42}\left[ L_{{\rm H\alpha }}\over {\rm erg.s}^{-1}\right] .  \label{29}
\end{equation}
which can be compared directly with the earlier calibrations by \citet{Dopita94,Kennicutt98,Panuzzo03}
(\emph{c.f.} equation \ref{2}). The agreement with these earlier works is satisfactory, given the 
different stellar models used here, and difference in other assumptions such as the IMF. Here we require 
a somewhat larger H$\alpha$ flux to deliver a given star formation rate. However, it 
should be emphasized that for other values of the molecular cloud dissipation timescale, equation \ref{27}
should be used, and these will require a greater star formation rate to produce a given H$\alpha$ flux.

\subsection{Star Formation Rates from the UV \& GALEX}
The GALEX satellite offers two wavebands centered at 1500\AA\ (FUV) and 2200\AA\ (NUV). We have folded the 
response function of these band passes with our predicted FUV SED. As can be seen from figure \ref{fig8}, 
the SED is unaffected by the ISM pressure. Instead, it is strongly influenced by the molecular cloud clearing
timescale. This is because, to first order, the molecular clouds act to obscure some fraction of the
exciting stars. Since, for any particular star, this obscuration is either non-existent or total, the
attenuation produced by the molecular clouds is effectively wavelength independent. The small changes in
the slope of the FUV spectrum, and in the relative intensity of particular spectral features in the 
UV, apparent in figure \ref{fig10}, are produced by age differences in the observable stellar populations.
In regions with long molecular cloud clearing timescales, the younger stars tend to be preferentially
obscured.

In real starburst galaxies, the FUV spectrum will also be obscured by the overlying foreground screen of
material. We will assume that this can be described by the attenuation curve given in figure \ref{fig5}
and table \ref{table3}. Here we are concerned only with the conversion of the absolute UV flux to star 
formation rate.

We find that for the model starbursts, the GALEX 1500\AA\ and 2200\AA\ bands provide nearly the same flux; 
$F(1500)/F(2200)=1.03\pm 0.015$, for all values of the molecular cloud dissipation timescale.
The $F(1500)$ flux can be fit, like
the H$\alpha$ flux in terms of the molecular cloud dissipation timescale. For all computed values of
$\tau$, this is given to better than $3\%$ accuracy by:
\begin{eqnarray*}
& log  \left[ {\rm L}_{1500} / {~\rm erg~s^{-1}Hz^{-1}} \right] =\\
&  28.62 +
log\left[SFR \over \rm {M_{\odot}~yr^{-1}}\right]
- log[2.3 +0.37  \left(\tau \over {\rm Myr}\right)]. \label{30}
\end{eqnarray*}

This flux accounts for the internal dust extinction (or, more properly, attenuation) of the \HII\ regions, 
but it does not account for any foreground screen attenuation.

The dependence of the flux on $\tau$ is weaker than for H$\alpha$ because the H$\alpha$ flux is produced
mostly in the first 4-5 Myr of a starburst, and so is very prone to obscuration by the placental molecular
clouds of the starburst itself. By contrast, the stellar UV flux at 1500\AA\ remains appreciable out to an age of 
tens of Myr and so is much less affected. Indeed, because of this the computed $F(1500)$ flux is somewhat 
sensitive to the assumed duration of the starburst, with older stars contributing significantly to the 
UV flux at this wavelength. For example, with $\tau=1$Myr, a $1.0~ \rm {M_{\odot}~yr^{-1}}$ starburst continued 
for $10^8$yr will give and absolute flux $F(1500)=1.58 \times 10^{28}{\rm erg~s}^{-1}~{\rm Hz}^{-1}$. However, 
if the duration of the starburst is only $10^7$yr the UV flux falls to 
$F(1500)=1.04 \times 10^{28}{\rm erg~s}^{-1}~{\rm Hz}^{-1}$.

For comparison with earlier calibrations, for a  $10^7$yr duration starburst with a molecular cloud
dissipation timescale, $\tau = 1$~Myr:

\begin{equation}
\left[SFR_{{\rm UV }}\over \rm {M_{\odot}~yr^{-1}}\right]
=0.96\times 10^{-28}\left[ L_{{1500}}\over{\rm erg~s}^{-1}~{\rm Hz}^{-1}\right] .  \label{31}
\end{equation}

Again, we require a larger observed $L_{{\rm 1500\AA}}$ flux to deliver a given star formation rate, but
larger values of the molecular cloud dissipation timescale require a greater star formation rate 
to produce a given $L_{{\rm 1500\AA}}$ flux.

\subsection{Star Formation Rates from the FIR}
In the far-IR, fluxes are affected by both the pressure and the molecular dissipation timescale, so the 
situation with respect to determining star formation rates is somewhat complex. If we were to choose a single 
wavelength for which the effect of the molecular dissipation timescale is largely compensated, then we would 
choose the IRAS 25\mum\ or the Spitzer Observatory MIPS 24\mum\ band. This is because at this wavelength, 
the reduction in the strength of the PAH bands for shorter molecular dissipation timescales is compensated by 
the increasing color temperature of the FIR re-emission continuum (see figures \ref{fig9}, \ref{fig10} and 
\ref{fig11}. This waveband probably gives the most reliable conversion of flux to star formation rate when 
$\tau$ is not known. In table \ref{table4} we give the computed luminosities through the MIPS 24\mum\ band and 
the IRAS 25\mum\ band for a star formation rate of $1.0~ \rm {M_{\odot}~yr^{-1}}$. With this star formation
rate these IR luminosities can be fit to an accuracy of $\pm 15$\%, which is probably sufficient for most 
astronomical purposes, by a simple function of pressure $log [{\rm L}_{25 \mu \rm m}/{\rm {L_{\odot}}}] = 
f(P/k)$ with $f(P/k)$ = 7.30, 7.69 and 8.00 (MIPS) and 7.66, 8.02 and 8.30 (IRAS) for the three pressures
$P/k=10^4, 10^6$ and $10^8$, respectively.

\begin{table}
  \caption{\label{table4} 
   Model Spizer \& IRAS 25\mum\ luminosities (given in terms of $log L/L_{\odot}$ units for a star formation
   rate of 1.0$M_{\odot}$yr$^{-1}$) as a function of both  $\tau$ and $P/k$.}
\begin{center}
  \begin{tabular}{lc|ccc}
  \hline
  \hline
   Waveband: & $\tau$ &$ P/k: $ & &\\ 
 & (Myr) & $10^4$ & $10^6$ & $10^7$\\     
 \hline
& & & & \\
   $log{L_{\rm MIPS(24)}}$ &  &  &  & \\
   & 1 & 7.395 & 7.676 & 7.952 \\
   & 2 & 7.331 & 7.676 & 7.976 \\
   & 4 & 7.288 & 7.683 & 8.000 \\
   & 8 & 7.271 & 7.694 & 8.022 \\
   & 16 & 7.267 & 7.706 & 8.039 \\
   & 32 & 7.268 & 7.714 & 8.049 \\  
 \hline
& & & & \\
   $log{L_{\rm IRAS(25)}}$ &  &  &  & \\
   & 1 & 7.739 & 8.003 & 8.245 \\
   & 2 & 7.682 & 8.006 & 8.269 \\
   & 4 & 7.645 & 8.014 & 8.295 \\
   & 8 & 7.631 & 8.026 & 8.324 \\
   & 16 & 7.630 & 8.038 & 8.337 \\
   & 32 & 7.630 & 8.046 & 8.348 \\  

     \hline
   \hline
  \end{tabular}
\end{center}
\end{table}

More usually, we are accustomed to using the conversion between far-IR luminosity $L_{\rm FIR}$, or the total
IR luminosity $L_{\rm IR}$. For the IRAS wavebands these are defined by \citep{Saunders96} as:
\begin{equation}
L_{\rm FIR} = 2.58L_{60\rm \mu m} + L_{100\rm \mu m}  \label{32}
\end{equation}
and
\begin{equation}
L_{\rm IR} = 13.48L_{12\rm \mu m}+ 5.16L_{25\rm \mu m} + 2.58L_{60\rm \mu m} + L_{100\rm \mu m}.  \label{33}
\end{equation}
With these definitions, we have computed the model fluxes for a star formation rate of 
$1.0~ \rm {M_{\odot}~yr^{-1}}$. These are given in table \ref{table5}. There is no simple fit to
these luminosities, owing to the sensitivity to $\tau$, particularly at the larger values of $P/k$.

When $\tau$ is long, and the pressure is high, then the majority of the stellar flux is re-radiated
in the far-IR, and $L_{\rm IR}$ becomes a good measure of the bolometric luminosity of the exciting
stars. This is particularly true at high pressure, since the dust temperature is then high enough that 
the emission beyond 100~\mum\ becomes unimportant. Put another way, the dust shell is virtually
complete around the stars and so the dust shell acts as an IR bolometer. For the most extreme case
computed ($\tau=32$Myr; $P/k=10^7$cm$^{-3}$K), $L_{\rm IR}$ provides a conversion to star formation
which can be compared with earlier estimates by \citet{Kennicutt98} and \citet{Panuzzo03}, (see equation
\ref{3}):

\begin{equation}
\left[SFR_{{\rm IR }}\over \rm {M_{\odot}~yr^{-1}}\right]
=4.2\times 10^{-44}\left[ L_{{\rm IR }} \over {\rm erg.s}^{-1}\right] .  \label{34}
\end{equation}

This agrees well with these earlier estimates. However, in general the dust shell will be more 
incomplete, and a greater fraction of luminosity is emitted longward of the 100~\mum\ band, so that
the numerical factor in equation \ref{34} will be larger. From table \ref{table4} we find that it
can range up to $11 \times 10^{-44}$, so that star formation estimates using equation
\ref{3} may underestimate the star formation rate by a factor of over two in certain circumstances.

\begin{table}
  \caption{\label{table5} 
   Model IRAS IR and FIR luminosities (given in terms of $log L/L_{\odot}$ units for a star formation
   rate of 1.0$M_{\odot}$yr$^{-1}$) as a function of $\tau$ and $P/k$.}
\begin{center}
  \begin{tabular}{lc|ccc}
  \hline
  \hline
   Waveband: & $\tau$ &$ P/k: $ & &\\ 
 & (Myr) & $10^4$ & $10^6$ & $10^7$\\     
 \hline
& & & & \\
   $log{L_{\rm FIR}}$ &  &  &  & \\
  & 1 & 9.131 & 9.164 & 9.063 \\
  & 2 & 9.111 & 9.200 & 9.155 \\
   & 4 & 9.096 & 9.263 & 9.261 \\
   & 8 & 9.103 & 9.329 & 9.342 \\
   & 16 & 9.116 & 9.380 & 9.450 \\
   & 32 & 9.127 & 9.405 & 9.496 \\
\hline
& & & & \\
   $log{L_{\rm IR}}$ &  &  &  & \\
   & 1 & 9.375 & 9.444 & 9.471 \\
   & 2 & 9.379 & 9.525 & 9.536 \\
   & 4 & 9.402 & 9.556 & 9.616 \\
   & 8 & 9.443 & 9.626 & 9.686 \\
   & 16 & 9.478 & 9.681 & 9.758 \\
   & 32 & 9.501 & 9.707 & 9.793 \\  
   \hline
   \hline
  \end{tabular}
\end{center}
\end{table}

\section{Conclusions}
In this paper we have attempted to provide a self-consistent theoretical modelling of the pan-spectral 
energy distribution of starbursts within the context of a simplified one dimensional evolutionary
model of the component \HII\ regions it contains. Although this modelling is rather more sophisticated than 
many models presented hitherto, it still suffers from a number of important limitations which we feel should
imply a \emph{caveat emptor} to the hopeful observationalist who would like to use the results presented here.
Enumerating these limitations:
\begin{itemize}
\item{The clusters are treated as having a single characteristic mass; $10^4 M_{\odot}$, rather than being 
distributed over all possible masses}.
\item{The one-dimensional model \HII\ evolution is oversimplified, as it does not take into account mixing and
dynamical instabilities which certainly will occur in two dimensions, nor does it take account of the intrinsic
inhomogeneities in the surrounding ISM. These effects have been incorporated in a `correction factor' which
reduces the efficiency of the stellar winds in inflating the surrounding bubble of ionized gas, and its magnitude
is chosen so as to provide ionization parameters $\cal U$ in the range observed. Since $\cal U$ also determines
the dust temperature in the surrounding shell, this choice of the correction also ensures that our computed 
dust temperatures in the molecular shells are more or less correct.}
\item{We appear to be missing a population of compact and ultra-compact \HII\ regions around single OB stars
or small OB clusters. These will provide an excess of flux in the $10-30$\mum band, and would allow a better
match to the observed SEDs of starbursts and to the observed color:color diagrams.}
\item{The dynamical evolution of the central cluster is not taken into account. This will alter the geometry 
of the stars with respect to the molecular gas, and would lead to different dust temperature distributions and
different effective covering factors of the molecular gas, $f$, as a function of time. This effect will become 
more important at high $P/k$ because the \HII\ region expansion `stalls' at an earlier phase of its evolution.}
\item{Collective effects are not taken into account. That is to say, we have not considered mutual overlap
of \HII\ regions along the line of sight, such as certainly occurs in very dense starbursts such as Arp220.
The treatment of such cases would require the SEDs computed here to be input as source functions into a 
complex dust radiative transfer code such as used by \citet{Popescu00}.}
\item{The diffuse component of the dust has only been included in respect of the attenuation of stellar light. 
The emission from this dust, which is expected to become dominant beyond 100\mum\ \citep{Popescu00}, is not
considered. Furthermore, old stars, that is to say stars older than $10^8$ years old, have not been included. 
As well as providing an additional component of direct light in the visual/NIR spectral range, these stars are 
also important in powering the diffuse dust emission component.}
\item{Observations of starburst galaxies \emph{e.g.} \citet{Meurer99,Calzetti01} show that in the visible 
and UV, much of the wavelength-dependent attenuation by dust can be understood as due to a foreground
diffuse screen of material. Much of this dust would have to lie outside the main star-forming region.
In our models we do not take this material into account, although we have provided the expected 
attenuation properties of this material in table \ref{table3}. This material may well provide an
important contribution to the far-IR dust continuum beyond 100\mum, and would again have to be accounted 
for within a full galaxian dust radiative transfer code.}
\end{itemize}
Despite these limitations, these models may well prove useful as providing SED templates for 
individual star formation regions within normal galaxies or for starburst galaxies that do not
contain an important old star population, or in which there is no extended diffuse dust component. 
For this reason, we are making available electronically both the computed SEDs, the attenuation functions and 
the filter transmissions (sampled at the same frequencies) for the various space missions referred to in 
this paper. These are available in the form of tab-delimited spreadsheets in the electronic version of the 
journal.

Despite the limitations listed above, this work provides a number of important advances:
\begin{itemize}
\item{We have demonstrated how the dust attenuation law of \citet{Calzetti01} for starburst galaxies can arise
as a natural consequence of turbulence on photochemistry in the diffuse phases of the ISM. In an earlier
paper, \citet{Fischera03} showed how the large values of $R_{\rm V}$ observed in starburst galaxies arise naturally
from the log-normal column density distribution imposed on the diffuse ISM by turbulent processes. In that work,
the observed absence of a 2200\AA\ dust absorption feature remained unexplained. Here we have shown how, if most
of the interstellar carbon is locked up into PAHs rather than graphitic or amorphous organic grains, then the
photodissociation of the PAHs in the strong UV radiation fields of starburst galaxies can explain the weakness
of the 2200\AA\ feature in absorption, whilst preserving the strong IR PAH emission features originating from the
photodissociation regions of molecular clouds.}
\item{ We have demonstrated that two ISM parameters control the form of the SEDs; the pressure in the diffuse phase
of the ISM (or, equivalently, the density), and the molecular cloud dissipation timescale. Long dissipation 
timescales for the molecular clouds enhance the PAH IR emission features by increasing the molecular 
cloud covering factors, but serve to obscure the central clusters in the visible and the UV. High pressures 
put the dense dusty gas closer to the exciting stars giving higher stochastic dust grain temperatures.}
\i.epstem{ We have demonstrated that the models are capable of describing the SED of well-observed starbursts over more 
than three decades of frequency, and we have shown that bolometrically-based star formation rate estimates
and foreground dust attenuations can be easily derived from this fitting.}
\item{We have shown that the models reproduce the observed IRAS colors and range of colors for starburst galaxies,
and we have provided diagnostic color:color plots for the Spitzer Space Observatory MIPS and IRAC instruments
which offer the potential to solve for the two controlling parameters of the SED.}
\item{We have provided fitting formulae or tables to enable the conversion of observed fluxes to star formation 
rates in the UV (GALEX), at optical wavelengths (H$\alpha$), and in the IR (IRAS or the Spitzer Space Observatory)
We have shown that the 25\mum\ fluxes are particularly valuable as star formation indicators since they depend 
little on the molecular gas dissipation timescale, and only weakly upon the ISM pressure. }
\end{itemize} 

Notwithstanding the limitations of these models, we hope that these results have been presented in such a way as 
to be useful for observers who are seeking both to understand star formation in galaxies, and in the Universe 
at large.

\begin{acknowledgements}
M. Dopita acknowledges the support of both the Australian National University and of
the Australian Research Council (ARC) through his ARC Australian Federation Fellowship. 
He would also like to thank Huub R\"ottgering for organising his vitit to Sterrewacht Leiden
where this work was started.
M. Dopita, R. Sutherland \& J. Fishera recognize the financial support of the ARC through 
Discovery project grant DP0208445. M. Reuland acknowledges the support of a travel grant
through M. Dopita's  ARC Australian Federation Fellowship Support Grant. 
L. Kewley is supported by a Harvard-Smithsonian CfA Fellowship. 
This research has made use of the NASA/IPAC Extragalactic Database (NED) which is operated by 
the Jet Propulsion Laboratory, California Institute of Technology, under contract with the National 
Aeronautics and Space Administration. Finally, all of us wish to thank the anonymous referee whose careful 
reading of the manuscript, and whose helpful suggestions has led to a much clearer end product.

\end{acknowledgements}


\begin{thebibliography}

\bibitem[Adelberger \& Steidel(2000)]{Adelberger00}  Adelberger, K.~L., \&
Steidel, C.~C. 2000, ApJ, 544, 218

\bibitem[Allain, Leach \& Sedlmayr(1996a)]{Allain96a} Allain, T., Leach, S. \&
Sedlmayr, E. 1996a,  A\&A 305, 602

\bibitem[Allain, Leach \& Sedlmayr(1996b)]{Allain96b} Allain, T., Leach, S. \&
Sedlmayr, E. 1996b,  A\&A 305, 616

\bibitem[Allen(1976)]{Allen76apj} Allen, D.~A.\ 1976, \apj, 207, 367 

\bibitem[Allende Prieto et al.(2001)]{Allende01} Allende Prieto, C.,
Lambert, D.~L. \& Asplund, M. 2001, \apjl, 556, L63

\bibitem[Allende Prieto et al.(2002)]{Allende02} Allende Prieto, C.,
Lambert, D.~L. \& Asplund, M. 2001, \apjl, (in press)

\bibitem[Allamandola, Hudgins, \& Sandford(1999)]{Allamandola99} 
Allamandola, L.~J., Hudgins, D.~M., \& Sandford, S.~A.\ 1999, \apjl, 511, 
L115

\bibitem[Asplund (2000)]{Asplund00a} Asplund, M., Nordlund, A.,
Trampedach, R. \& Stein, R.~F. 2000, A\&A, 359, 743

\bibitem[Asplund et al.(2000)]{Asplund00b} Asplund, M. 2001, A\&A, 359, 755.

\bibitem[Barnes \& Hernquist(1996)]{Barnes96} Barnes, J.~E., Hernquist, L.
1996, \apj, 471, 115

\bibitem[Barton, Geller, \& Kenyon(2000)]{Barton00} Barton, B.~J.,
Geller, M.~J., \& Kenyon, S.~J. 2000, \apj, 530, 660

\bibitem[Bell(2003)]{Bell03} Bell, E.~F. 2003, \apj, 586, 794

\bibitem[Bell \& Kennicutt(2001)]{Bell01}  Bell, E.~F,. \& Kennicutt, R.~C.
Jr. 2001, ApJ, 548, 681

\bibitem[Benford(1999)]{Benford99thesis} Benford, D.~J.\ 1999, 
unpublished Ph.D.~Thesis  

\bibitem[Blain et al.(2004)]{Blain04}  Blain, A.~W., Chapman, S.~C., Smail, I., 
\& Ivison, R. 2004, astro-ph/0404438

\bibitem[Bohlin et al.(1978)]{Bohlin78}Bohlin, R.~C., Savage, B.~D. 
\& Drake, J.~F. 1979, \apj, 224, 132

\bibitem[Bottorff et al.(1998)]{Bottorff98} Bottorff, M., LaMothe, J.,
Momjian, E., Verner, E., Vinkovi«c, D., \& Ferland, G. 1998, PASP, 110, 1040.

\bibitem[Boulanger et al.(1998)]{Boulanger98}  Boulanger, F., Boissel, P., 
Cesarsky, D., \& Ryter,C. 1998, A\&A, 339, 194

\bibitem[Bressan, Silva \& Granato(2002)]{Bressan02}  Bressan, A., 
Silva, L.  \&  Granato, G.~L. 2002, \aap, 392, 377.

\bibitem[Buat et al.(1999)]{Buat99}  Buat, V., Donas, J., Milliard, B. \& 
Xu, C. 1999, \aap, 352, 371

\bibitem[Burton et al.(2000)]{Burton00}Burton, M.~G., Ashley, M.~C.~B., Marks, R.~D.,
Schinckel, A.~E., Storey, J.~W.~V., Fowler, A., Merrill, M., Sharp, N., Gatley, I.
Harper, D.~A., Loewenstein, R.~F., Mrozek, F., Jackson, J.~M.\& Kraemer, K.~E.,
2000, \apj, 542, 359

\bibitem[Bushouse(1987)]{Bushouse87} Bushouse, H. A. 1987, \apj, 320, 49

\bibitem[Calzetti(2001)]{Calzetti01}  Calzetti, D. 2001, PASP, 113, 1449.

%\bibitem[Calzetti et al.(1994)]{Calzetti94}Calzetti, D. Kinney, A.~L.
%\& Storchi-Bergmann, T. 1994, \apj, 429, 582

\bibitem[Cardelli, Clayton, \& Mathis(1989)]{Cardelli89apj} Cardelli,
J.A., Clayton, G.C.\ \& Mathis, J.S. 1989, \apj, 345, 245
 
\bibitem[Cho \& Vishniac(2000b)]{ChoVish00b} Cho, J.~\& Vishniac, 
E.~T.\ 2000b, \apj, 539, 273


\bibitem[Clements et al.(2002)]{Clements2002} Clements,ÊD.~L., McDowell,ÊJ.~C., 
Shaked,ÊS., Baker,ÊA.~C., Borne,ÊK., Colina,ÊL., Lamb,ÊS.~A. \& Mundell,ÊC 2002, 
\apj, 581, 974.

%\bibitem[Colina, Sparks, \& Macchetto(1991)]{Colina91}  Colina, L., Sparks,
%W. B., Macchetto, F. 1991, \apj, 370, 102

%\bibitem[Condon et al.(1982)]{Condon82}  Condon, J. J., Condon, M. A.,
%Gisler, G., \& Puschell, J. J. 1982, \apj, 252, 102

\bibitem[Condon(1992)]{Condon92}  Condon, J.~J. 1992, \araa, 30, 575

\bibitem[Condon \& Dressel(1978)]{Condon78} Condon, J.~J. \& Dressel, L.
L.
1978, \apj, 221, 456

\bibitem[Condon {\it et al.}(1982)]{Condon82} Condon, J.~J., Condon, M.
A., Gisler, G., \& Puschell, J.~J. 1982, \apj, 252, 102

%\bibitem[Condon et al.(1996)]{Condon96}  Condon, J.~J., Helou, G., Sanders,
%D.~B., \& Soifer, B.~T. 1996, \apjs, 103, 81

%\bibitem[Condon et al.(1998)]{Condon1998}  Condon, J.~J., Cotton, W.~D.,
%Greisen, E.~W., Yin, Q.~F.; Perley, R.~A.; Taylor, G.~B., \& 
%Broderick, J.~J. 1998, \aj, 115, 1693

%\bibitem[Cram et al.(1998)]{Cram98}  Cram, L., Hopkins, A., Mobasher, B., \&
%Rowan-Robinson, M. 1998, \apj, 507, 155

\bibitem[Dale et  al.(2001)]{Dale01} Dale, D.~A., Helou, G., Contursi, A., 
Silberman, N.~A., \&  Kolhatkar, S. 2001, \apj, 549, 215.

\bibitem[Dale \& Helou(2002)]{Dale02} Dale, D.~A., \& Helou, G. 2002, \apj, 576, 159.

\bibitem[de Vaucouleurs et al.(1991)]{deVaucouleurs91cat} de Vaucouleurs, G., 
de Vaucouleurs, A., Corwin, H.~G., Buta, R.~J., Paturel, G., \& Fouque, P.\ 1991, 
Volumes 1-3,  Springer-Verlag:Berlin 

\bibitem[Donzelli \& Pastoriza(1997)]{Donzelli97} Donzelli, C.~J. \&
Pastoriza, M.~G. 1997, \apjs, 111, 118

\bibitem[Dopita  et al.(2004)]{Dopita04}Dopita,ÊM.ÊA., Fischera, J., Groves,ÊB., 
Sutherland,ÊR.~S., Kewley,ÊL.~J., Tuffs, R., Popescu, C. \& Leitherer, C. 2004, 
in Proc. IAU Symp 222, ed. T. Storchi-Bergmann, (CUP: Cambridge), in press.

\bibitem[Dopita et al.(1998)]{Dopita98}  Dopita M.~A., Heisler, C.~A., Lumsden, S. \& 
Bailey, J. 1998, \apj, 498, 570

\bibitem[Dopita et al.(2000)]{Dopita00}
Dopita, M.~A., Kewley, L.~J., Heisler, C.~A., \& Sutherland, R.~S.\ 2000,
\apj, 542, 224

\bibitem[Dopita  et al.(2002)]{Dopita02b}Dopita, M.~A., Groves, B.~A., Sutherland, R.~S., 
Binette, L. \& Cecil, G. 2002, \apj, 572, 753

\bibitem[Dopita et al.(2003)]{Dopita03}Dopita,ÊM.~A., Groves,ÊB~ÊA., Sutherland,ÊR~ÊS., 
\& Kewley,ÊL.~J.  2003, \apj, 583, 727

\bibitem[Dopita  et al.(2002)]{Dopita02}Dopita, M.~A., Pereira, M. Kewley, 
L.~J. \& Capaccioli, M. 2002, \apjs, 143, 47

\bibitem[Dopita \& Ryder(1994)]{Dopita94}  Dopita, M.~A. \& Ryder,S.~D. 1994,
ApJ, 430, 163.

\bibitem[Dopita \& Sutherland(2003)]{Dopandsuth03}  Dopita, M.~A. \& Sutherland, R.~S.
2003, ``Diffuse  Matter in  the Universe", (Springer: Heidelberg).

\bibitem[Draine \& Anderson(1985)]{Draine85} Draine, B.~T.~\& 
Anderson, N.\ 1985, \apj, 292, 494 

\bibitem[Draine \& Li(2001)]{Draine01} Draine, B.~T.~\& Li, A\
2001, \apj, 551, 807

\bibitem[Dwek(1986)]{Dwek86} Dwek, E. 1986, \apj, 302, 363

\bibitem[Dwek(1998)]{Dwek98} Dwek, E., Arendt, R.~G., Hauser, M.~G., Fixsen, D., 
Kelsall, T., Leisawitz, D., Pei, Y.~C., Wright, E.~L., Mather, J.~C.,Moseley, S.~H., 
Odegard, N., Shafer, R., Silverberg, R.~F.., \& Wiland, J.~L. 1998, \apj, 508, 106

\bibitem[Duley(1973)]{Duley73} Duley, W.~W.\ 1973, \apss, 23, 43 


\bibitem[Dunne et al.(2000)]{Dunne00mnras} Dunne, L., Eales, S., 
Edmunds, M., Ivison, R., Alexander, P., \& Clements, D.~L.\ 2000, \mnras, 
315, 115 

\bibitem[Dunne \& Eales(2001)]{Dunne01mnras} Dunne, L.~\& Eales, S.~A.\ 2001, 
\mnras, 327, 697 


\bibitem[Fischera et al.(2003)]{Fischera03} Fischera,ÊJ., Dopita,ÊM.~A., 
\& Sutherland,ÊR.~S.,  2003, \apjl, 599, L21

\bibitem[Gallego et al.(1995)]{Gallego95}  Gallego, J., Zamorano, J.,
Arag\'{o}n-Salamanca, A., \& Rego, M. 1995, ApJ, 455, L1.

\bibitem[Garc\'{\i}a-Segura et  al.(1995)]{Garcia95}Garc\'{\i}a-Segura, G. 
\& Mac Low,  M.-M.  1995, \apj, 455, 155

%\bibitem[Genzel et al.(1998)]{Genzel98}  Genzel, R., Lutz, D., Sturm, E.,
%Egami, E., Kunze, D., Moorwood, A.~F.~M., Rigopoulou, D., Spoon, H.~W.~W.,
%Sternberg, A., Tacconi-Garman, L. E., Tacconi, L., \& Thatte, N. 1998, \apj,
%498, 579

\bibitem[Gispert et al.(2000)]{Gispert00}Gispert, R., Lagache, G., \& 
Puget, J.~L. 2000, \aa 360, 1

\bibitem[Goldader et al.(1997)]{Goldader97}  Goldader, J.~D., Joseph, R.~D.,
Doyon, R., \& Sanders, D.~B. 1997, \apj, 474, 104

%\bibitem[Goldader et al.(1997b)]{Goldader97b}  Goldader, J. D., Joseph, R.
%D., Doyon, R., \& Sanders, D. B. 1997b, \apjs, 108, 449

\bibitem[Granato et al.(2000)]{Granato00} Granato, G.~L., Lacey, C.~G., Silva, L,  
Bressan, A., Baugh, C.~M.,  Cole, S. \&  Frenk, C.~S. 2000, \apj, 542, 710

\bibitem[Greenberg \& Hong(1974)]{Greenberg74} Greenberg, J.~M.~\& 
Hong, S.-S.\ 1974, IAU Symp.~ 60: Galactic Radio Astronomy, 60, 155

\bibitem[Groves et al.(2003)]{Groves03} Groves, B.~A., Cho, J., Dopita, M., 
\& Lazarian, A., 2003, PASA, 20, 252

\bibitem[Guhathakurta \& Draine(1989)]{Guhatha89} Guhathakurta, 
P.~\& Draine, B.~T.\ 1989, \apj, 345, 230

%\bibitem[de Grijp et al.(1992)]{deGrijp92}  de Grijp, M.~H.~K., Keel, W.~C.,
%Miley, G.~K., Goudfrooij, P., \& Lub, J. 1992, \aap, 96, 389

\bibitem[Haas et al.(1998)]{Haas98} Haas,ÊM., Lemke,ÊD., Stickel,ÊM., 
Hippelein,ÊH., Kunkel,ÊM., Herbstmeier,ÊU., \& Mattila,ÊK. 1998, \aap, 338, L33

\bibitem[Hippelein et al.(2003)]{Hippelein03} Hippelein,ÊH., Haas,ÊM., 
Tuffs,ÊR.~J., Lemke,ÊD., Stickel,ÊM., Klaas,ÊU., \& V\"olk,ÊH.ÊJ.  2003, \aap, 407, 137

\bibitem[Hillier \& Miller(1998)]{Hillier98} Hillier, D.~J. \& Miller, D. L. 
1998, \apj, 496, 407

\bibitem[Hirashita \& Ferrara(2002)]{Hirashita02} Hirashita, H.,  
\& Ferrara, A. 2002, \mnras, 337, 921.

\bibitem[Hirashita, Buat \& Inoue(2003)]{Hirashita03} Hirashita, H.,  
Buat, V. \& Inoue,  A.~K. 2003, \aap, 410, 83.

\bibitem[Hummel(1980)]{Hummel80} Hummel, E. 1980, \aap, 89, L1

\bibitem[Hummel {\it et al.}(1990)]{Hummel90} Hummel, E., van der Hulst,
J.~M., Kennicutt, R.~C., \& Keel, W. C. 1990, \aap, 236, 333

%\bibitem[Igl\'{e}sias-Paramo, \& Vilchez(1997)]{Iglesias97}  
%Igl\'{e}sias-Paramo, J., \& Vilchez, J.M. 1997\apj, 479, 190

%\bibitem[Igl\'{e}sias-Paramo \& Vilchez(2001)]{Iglesias01}  
%Igl\'{e}sias-Paramo, J,. \& Vilchez, J. M. 2001, \apj, 550, 204

\bibitem[Ikebe et al.(2000)]{Ikebe00}  Ikebe, Y., Leighly, K., Tanaka, Y.,
Nakagawa, T., Terashima, Y., \& Komossa, S. 2000, \mnras, 316, 433

\bibitem[Inoue, Hirashita, \& Kamaya(2000)]{Inoue00}  Inoue, A.~K.,
Hirashita, H., \& Kamaya, H. 2000, PASJ, 52, 539.

\bibitem[Inoue, Hirashita \& Kamaya(2001)]{Inoue01a}  Inoue, A.~K.,
Hirashita, H., \& Kamaya, H. 2001, \apj, 555, 613

\bibitem[Inoue(2001)]{Inoue01b}  Inoue, A.~K., 2001, \aj, 122, 1788

\bibitem[Inoue(2002)]{Inoue02} Inoue, A.~K. 2002, \apj, 570, L97

\bibitem[Jenkins(1983)]{Jenkins83} Jenkins, E.~B.\ 1983

\bibitem[Jenkins, Jura \& Loewenstein(1987)]{Jenkins87}  
Jenkins, E.~B., Jura, M., \& Loewenstein, M. 1987, \apj, 270, 88

\bibitem[Jones \& Bland-Hawthorn(2001)]{Jones01}  Jones, D.~H. \&
Bland-Hawthorn, J. 2001, ApJ, 550, 593

\bibitem[Jones, Tielens \& Hollenbach(1996)]{Jones96} Jones,A.~P.,
Tielens, A.~G.~G.~M. \& Hollenbach, D.~J. 1996, \apj, 469, 740

\bibitem[Keel {\it et al.}(1985)]{Keel85} Keel, W.~C., Kennicutt, R.~C.,
Hummel, E., \& van der Hulst, J.~M. 1985, \aj, 90, 708

\bibitem[Kennicutt(1998)]{Kennicutt98}  Kennicutt, R.~C. Jr. 1998,
\araa, 36, 189.

\bibitem[Kennicutt {\it et al.}(1987)]{Kennicutt87} Kennicutt, R.~C.,
Keel, W.~C., van der Hulst, J.~M.,Hummel, E. \& Roettiger, K.~A. 1987, 
\aj, 95, 5

\bibitem[Kewley et al.(2000)]{Kewley00}  Kewley, L.~J., Heisler, C.~A.,
Dopita, M.~A., Sutherland, R.~S., Norris, R., Reynolds, J., Lumsden, S. 
2000, \apj, 530, 704

\bibitem[Kewley et al.(2001)]{Kewley01}  Kewley, L.~J., Dopita, M.~A.,
Sutherland, R.~S., Heisler, C. ~A., \& Trevena, J. 2001, \apj, 556, 121

\bibitem[Kewley et al.(2001b)]{Kewley01b}  Kewley, L.~J., Heisler, C.~A.,
Dopita, M.~A., Lumsden, S. 2001b, \apjs, 132, 37

\bibitem[Kewley \& Dopita(2002)]{Kewley02}  Kewley, L.~J., \& Dopita, M.~A.
2002, Kewley, L.~J. \& Dopita, M.~A. 2002, \apjs, 142, 35

\bibitem[Kim et al.(1995)]{Kim95}  Kim, D.-C., Sanders, D.~B. Veilleux, S.,
Mazzarella, J.~M., Soifer, B.~T. 1995, \apjs, 98, 129

\bibitem[Klaas et al.(1997)]{Klaas97aa} Klaas,U., Haas, M., Heinrichsen, I., 
\& Schulz, B.\ 1997, \aap, 325, L21

\bibitem[Klaas et al.(2001)]{Klaas01aa} Klaas, U., Haas,ÊM., 
M\"uller,ÊS.~A.~H., Chini,ÊR., Schulz,ÊB., Coulson,ÊI., Hippelein,ÊH, 
Wilke,ÊK., Albrecht,ÊM., \& Lemke,ÊD. 2001, \aap, 379, 823

\bibitem[Kurucz(1992)]{Kurucz92}  Kurucz, R.~L. 1992, in IAU Symp. 149, 
"The Stellar Populations of Galaxies", ed. B. Barbuy \& A. Renzini
(Dordrecht:Kluwer), 225

\bibitem[Lambas et al.(2003)]{Lambas03} Lambas, D.~G., Tissera, P.~B., 
Sol Alonso, M., \& Coldwell, G. 2003, \mnras, in press (astro-ph/0212222)

\bibitem[Larson \& Tinsley(1978)]{Larson78} Larson, R.~B. \& 
Tinsley, B.~M. 1978, \apj, 219, 46

\bibitem[Laurent et al.(2000)]{Laurent00}  Laurent, O., Mirabel, I.~F., 
Charmandaris, V., Gallais, P., Madden, S.~C., Sauvage, M., Vigroux, L. 
\& Cesarsky, C. 2000, A\&A, 359, 887.

\bibitem[L\'eger \& Puget(1984)]{Leger84} L\'eger, A. \& Puget, J.~L. 1984, 
\aap, 137, 5

%\bibitem[Leitherer \& Heckman(1995)]{Leitherer95}  Leitherer, C., \& Heckman,
%T. M. 1995, \apjs, 96, 9

\bibitem[Leitherer et al.(1999)]{Leitherer99}  Leitherer, C., Schaerer, D.,
Goldader, J.~D., Delgado, R.~M. Gonz\'{a}lez, R.~C., Kune, D.~F., de Mello,
Du\'{i}lia F., Devost, D., \& Heckman, T.~M. 1999, \apjs, 123, 3

\bibitem[Lejeune, Cuisinier \& Buser(1997)]{Lejeune97}  Lejeune,
Th., Cuisinier, F. \& Buser, R. 1997, A\&AS, 125, 229

%\bibitem[Lisenfeld, V\"olk \& Xu(1996)]{Lisenfeld96}  Lisenfeld, U.,
%V\"olk, H.~J., \& Xu , C. 1996, \aa, 306, 677

\bibitem[Li \& Draine(2001)]{Li01} Li, A.~\& Draine, B.~T.
2001, \apj, 554, 778

%\bibitem[Lui \& Kennicutt(1995)]{Lui95}  Lui, T.~L. \& Kennicutt, R.~C.
%1995, \apj, 450, 547

\bibitem[Madau et al.(1996)]{Madau96}  Madau, P. , Ferguson, H.~V.~C.,
Dickinson, M.~E., Giavalisco, M., Steidel, C.~C., \& Fruchter, A. 1996, MNRAS,
283, 1388.

\bibitem[Madore(1986)]{Madore86} Madore, B.~F. 1986, in "Spectral Evolution
of Galaxies", ed. C. Chiosi \& A. Renzini (Dordrecht: Reidel), 97

\bibitem[Mathis, Rumpl, \& Nordsieck(1977)]{MRN77} Mathis,
J.~S., Rumpl, W., \& Nordsieck, K.~H.\ 1977, \apj, 217, 425 (MRN)

\bibitem[Mezger, Smith, \& Churchwell(1974)]{Mezger74} Mezger, P.~G., 
Smith,L.~F., \& Churchwell, E. 1974, \aap, 32, 269.

\bibitem[Meurer et al.(1997)]{Meurer97}  Meurer, G.~R., Heckman, T.~M., 
Lehnert, M.~D., Leitherer, C., \& Lowenthal, J. 1997, \aj, 114, 54

\bibitem[Meurer, Heckman \& Calzetti(1999)]{Meurer99}  Meurer, G.~R.,
Heckman, T.~M., \& Calzetti, D. 1999, ApJ, 521, 64.

\bibitem[Misiriotis et al.(2001)]{Misiriotis01} Misiriotis, A., Popescu, C.~C., 
Tuffs, R.~J., \& Kylafis, N.~D. 2000, \aap, 372, 775

\bibitem[Moorwood et al.(2000)]{Moorwood00}  Moorwood, A.~F.~M, van der Werf,
P.~P., Cuby, J.~G., \& Oliva, E. 2000, \aap, 362, 9

%\bibitem[Morganti et al.(1999)]{Morganti99}  Morganti, R., Tsvetanov, Z.~I.,
%Gallimore, J., \& Allen, M.~G. 1999, A\&AS, 137, 457

\bibitem[Natta \& Panagia(1976)]{Natta76} Natta, A. \& Panagia, N. 1976,
\aap, 50, 191

\bibitem[Naz\'e et al.(2001)]{Naze01}Naz\'e, Y.,  Chu, Y.-H., 
Points, S. ~D.,  Danforth, C.~W., Rosado, M., \&  Chen, C.-H.~R.  2001, \aj, 122, 921

\bibitem[Niklas, Klein \& Wielebinski (1997)]{Niklas97} Niklas, S., 
Klein, U. \& Wielebinski, R., 1997, \aap, 322, 19

%\bibitem[Norris et al.(1990)]{Norris90}  Norris, R.~P., Allen, D.~A., Sramek,
%R.A., Kesteven, M.~J. \& Troup, E.~R. 1990,\apj, 359, 291

%\bibitem[Norris et al.(1988)]{Norris88}  Norris, R.~P., Kesteven, M.~J.,
%Wellington, K.J. \& Batty, M.~J. 1988, \apjs, 67, 85

\bibitem[Oey(1996)]{Oey96}Oey, M. S. 1996, \apj, 467, 666

\bibitem[Oey \& Clarke(1997)]{Oey97} Oey, M.~S., \& Clarke, C.~J. 1997,
\mnras, 289, 570.

\bibitem[Oey \& Clarke(1998)]{Oey98} Oey, M.~S., \& Clarke, C.~J. 1998,
\aj, 115, 1543.

\bibitem[Panagia(1974)]{Panagia74} Panagia, N. 1974, ApJ, 192, 221

%\bibitem[Papdopoulos \& Allen(2000)]{Papadopoulos00}  Papadopoulos, P.~P.
%\& Allen, M.~L. 2000 \apj, 537, 631

\bibitem[Panuzzo et al.(2003)]{Panuzzo03} Panuzzo, P., Bressan, A., Granato, G.~L., 
Silva, L. \& Danese, L. 2003, \aap, 409, 99

\bibitem[Pascual et al.(2001)]{Pascual01}  Pascual, S., Gallego, J.,
Arag\'{o}n-Salamanca, A., \& Zamorano, J. 2001, \aap, 379, 798.

\bibitem[Pauldrach, Hoffman, \& Lennon(2001)]{Pauldrach01} Pauldrach, A.
W.~A., Hoffman, T.~L., \& Lennon, M. 2001, \aap, 374, 161

\bibitem[Peeters(2002)]{Peeters02} Peeters, E. 2002, Thesis, 
Rijksuniversiteit Groningen

\bibitem[Petrosian(1978)]{Petrosian78} Petrosian, A.~R., Saakian, K.~A.,
\& Khachikian, E~ E. 1978, Astrofizika, 14, 69

\bibitem[Petrosian et al.(2002)]{Petrosian02} Petrosian, A., McLean, B.,
Allen, R.~J., Leitherer, C., MacKenty, J., \& Nino, P. 2002, \aj, 123,
2280

\bibitem[Petrosian, Silk \& Field(1972)]{Petrosian72}  Petrosian, V., Silk,
J., \& Field, G.~B. 1972, \apj, 177, L69.

%\bibitem[Phillips, Charles, \& Baldwin(1983)]{Phillips83}  Phillips, M.~M.,
%Charles, P.~A., \& Baldwin, J.~A. 1983, \apj, 266, 485

\bibitem[Pierini et al.(2003)]{Pierini03}Pierini, D., Popescu, C.~C., Tuffs, R.~J., 
\& V\"olk, H.~J. 2003, \aap, 409, 907

\bibitem[Popescu et al.(2000)]{Popescu00}  Popescu, C.~C., Misiriotis, A.,
Kylafis, N.~D., Tuffs, R.~J., \& Fischera, J. 2000, \aap, 362, 138


\bibitem[Popescu et al.(2004)]{Popescu04} Popescu,ÊC.~C., Tuffs,ÊR.~J., 
Kylafis,ÊN.ÊD., \& Madore,ÊB.ÊF. 2004, \aap, 414, 45

\bibitem[Popescu et al.(2002)]{Popescu02} Popescu,ÊC.~C., Tuffs,ÊR.~J.,
V\"olk,ÊH.~J., Pierini,ÊD., \& Madore,ÊB.~F. 2002, \apj, 567, 221

\bibitem[Purcell(1976)]{Purcell76} Purcell, E.~M.\ 1976, \apj, 
206, 685

%\bibitem[Rafanelli \& Marziani(1992)]{Rafanelli92}  Rafanelli, P. \&
%Marziani, P. 1992, \aj, 103, 743

%\bibitem[Read \& Ponman(1998)]{Read98}  Read, A.~M. \& Ponman, T.~J. 1998, 
%\mnras, 297, 143

%\bibitem[Richter, Sackett, \& Sparke(1994)]{Richter94}  Richter, O.-G.,
%Sackett, P.~D., \& Sparke, L.~S. 1994, \aj, 107, 99

\bibitem[Risaliti et al.(2000)]{Risaliti00}  Risaliti, G., Gilli, R.,
Maiolino, R., \& Salvati, M. 2000, \aap, 357, 13

\bibitem[Rush, Malkan \& Spinoglio(1993)]{Rush93} Rush, B., 
Malkan, M.~A., \& Spinoglio, L. 1993, \apjs, 89, 1

\bibitem[Sarazin(1977)]{Sarazin77} Sarazin, C.~L., 1977, \apj, 211, 722.

\bibitem[Saunders \& Mirabel(1996)]{Saunders96} Saunders, D.~B., 
\& Mirabel, I.~F. 1996, \araa, 34, 749

\bibitem[Savage \& Sembach(1996)]{SavageS96} Savage, B.~D.~\&
Sembach, K.~R.\ 1996, \araa, 34, 279

\bibitem[Sawaki \& Yee(1998)]{Sawicki98}  Sawicki, M., \& Yee, H.~K.~C.
1998, AJ, 115, 1329

\bibitem[Schlegel, Finkbeiner, \& Davis(1998)]{Schlegel98apj}
Schlegel, D.~J., Finkbeiner, D.~P., \& Davis, M.\ 1998, \apj, 500, 525 

%\bibitem[Schweizer(1982)]{Schweizer82}  Schweizer, F. 1982, \apj, 252, 455

%\bibitem[Schulz et al.(1993)]{Schulz93}  Schulz, H., Fried, J.~W., 
Roser,S., \& Keel, W.~C. 1993, \aap, 277, 416

\bibitem[Sellgren(1984)]{Sellgren84} Sellgren, K.\ 1984, \apj, 277, 623

\bibitem[Sellgren, Werner, \& Dinerstein(1983)]{Sellgren83} 
Sellgren, K., Werner, M.~W., \& Dinerstein, H.~L.\ 1983, \apjl, 271, L13

\bibitem[Sersic \& Pastoriza(1967)]{Sersic67} Sersic, J.~L. \& 
Pastoriza, M. 1967, \pasp, 79, 152

\bibitem[Shields \& Kennicutt(1995)]{Shields95} Shields, J.~C. \& Kennicutt,
R.~C. 1995, ApJ, 454, 807.

%\bibitem[Shields \& Filippenko(1990)]{Shields90}  Shields, J.~C. \&
%Filippenko, A.~V. 1990, \aj, 100, 1034

\bibitem[Silva et al.(1998)]{Silva98} Silva, L., Granato, G.~L., Bressan, A., 
\& Danese, L. 1998, \apj, 509, 103.

\bibitem[Smith, Biermann, \& Mezger(1978)]{Smith78} Smith, L.~F., Biermann,
P. \& Mezger, P.~G. 1978, A\&A, 66, 65

\bibitem[Smith, Norris, \& Crowther(2002)]{Smith02} Smith, L.~J., Norris,
R.~P.~F., \& Crowther, P.~A. 2002, \mnras, 337, 1309

%\bibitem[Soifer et al.(1987)]{Soifer87}  Soifer, B.~T., Sanders, D.~B.,
%Madore, B.~F., Neugebauer, G., Danielson, G.~E., Elias, J.~H., Lonsdale, C.
%J., \& Rice, W. L. 1987, \apj, 320, 238

\bibitem[Spergel et al.(2003)]{Spergel03apj} Spergel, D.~N., et al.\
2003, \apjs, 148, 175

\bibitem[Spinoglio et al.(1995)]{Spinoglio95apj} Spinoglio, L.,
Malkan, M.~A., Rush, B., Carrasco, L., \& Recillas-Cruz, E.\ 1995,
\apj, 453, 616

\bibitem[Spoon et al.(2004)]{Spoon04aa} Spoon, H.~W.~W., 
Moorwood, A.~F.~M., Lutz, D., Tielens, A.~G.~G.~M., Siebenmorgen, R., \& 
Keane, J.~V.\ 2004, \aap, 414, 873 

%\bibitem[Stockton(1974)]{Stockton74}  Stockton, A. 1974, \apj, 190, 47

%\bibitem[Strauss et al.(1992)]{Strauss92}  Strauss, M. A., Huchra, J. P.,
%Davis, M., Yahil, A., Fisher, K.~B., \& Tonry, J. 1992, \apjs, 83, 29

\bibitem[Sulentic(1976)]{Sulentic76} Sulentic, J.~W. 1976, \apjs, 32, 171

%\bibitem[Sullivan et al.(2001)]{Sullivan01}  Sullivan, M., Mobasher, B.,
%Chan, B., Cram, L., Ellis, R., Treyer, M,. \& Hopkins, A. 2001, ApJ, 558, 72

\bibitem[Sutherland \& Dopita(1993)]{Sutherland93} Sutherland, R.~S. \& 
Dopita, M.~A. 1993, \apjs, 88, 253

\bibitem[Tagaki, Vansevicius \& Arimoto(2003)]{Takagi03a} Tagaki, T., 
Vansevicius, V., \& Arimoto, N. 2003, \pasj, 55, 385.

\bibitem[Tagaki, Arimoto \& Hanami(2003)]{Takagi03b} Tagaki, T., 
Arimoto, N. \& Hanami, H. 2003, \mnras, 340, 813

\bibitem[Takeuchi et al.(2003)]{Takeuchi03} Takeuchi, T.~T., Hirashita, H., 
Ishii, T~ T., Hunt, L.~K. \& Ferrara, A. 2003, \mnras, 343, 839

%\bibitem[Telesco, Wolstencroft, \& Done(1988)]{Telesco88}  Telesco, C.~M.,
%Wolstencroft, R.~D., \& Done, C. 1988, \apj, 329, 174

\bibitem[Tonry et al.(2003)]{Tonry03apj} Tonry, J.~L., et al.\ 
2003, \apj, 594, 1 

\bibitem[Trager et al.(1997)]{Trager97}  Trager, S.~C., Faber, S.~M.,
Dressler, A., \& Oemler, A. Jr. 1997, ApJ, 485, 92

\bibitem[Tresse \& Maddox(1998)]{Tresse98}  Tresse, L., \& Maddox, S.~J.
1998, ApJ, 495, 691

%\bibitem[Turner et al.(1997)]{Turner97}  Turner, T.~J., George, I.~M.,
%Nandra, K., \& Mushotsky, R.~F. 1997, \apjs, 113, 23

\bibitem[Tuffs \& Popescu(2003)]{Tuffs03}Tuffs, R.~J. \& Popescu, C.~C. 
2003, ESA SP-511, Eds. C. Gry, S. Peschke, J. Matagne, P. Garcia-Lario, 
R. Lorente, \& A. Salama. (ESA: Noordwijk), p. 239

\bibitem[Tuffs et al.(2004)]{Tuffs04} Tuffs, R. J., Popescu, C.~C., 
V\"olk, H. J., Kylafis, N.~D. \& Dopita, M.~A. 2004, \aap, 419, 821

%\bibitem[Vacca \& Conti(1992)]{Vacca92}  Vacca, W. D. \& Conti, P. S. 1992, 
%\apj, 401, 543

\bibitem[Van Kerckhoven et al.(2000)]{VanKerckhoven00}  Van Kerckhoven, C., 
Hony, S., Peeters, E., Tielens, A.~G.~G.~M., Allamandola. L.~J., Hudgins, D.~M., 
Cox, P., Roelfsema, P.~R., Voors, R.~H.~M., Waelkens, C., Waters, L.~B.~F~.M. 
\& Wesselius, P.~R. 2000, \aap, 357, 1013

%\bibitem[van den Broek et al.(1991)]{vandenBroek91}  van den Broek, A. C.,
%van Driel, W., de Jong, W., Lub, J., de Grijp, M. H. K., \& Goudfrooij, P.
%1991, A\&AS, 91, 61

\bibitem[Veilleux et al.(1995)]{Veilleux95}  Veilleux, S., Kim, D.-C.,
Sanders, D.~B., Mazzarella, J.~M., \& Soifer, B.~T. 1995, \apjs, 98, 171

\bibitem[Verstraete al.(2001)]{Verstraete01} Verstraete, L., Pech, C., 
Moutou, C., Sellgren, K., Wright, C.~M., Giard, M., L\'{e}ger, A., 
Timmermann, R. \& Drapatz, S. 2001, \aap, 372, 981.

\bibitem[Vorontsov-Velyaminov(1959)]{Vorontsov59} Vorontsov-Velyaminov, B.
A. 1959, Atlas and Catalogue of Interacting Galaxies, Sternberg Institute
(Moscow: Moscow State University)

%\bibitem[Wang \& Heckman (1996)]{Wang96}  Wang, B., \& Heckman, T.~M. 1996
% \apj, 457, 645

\bibitem[Weaver et al.(1977)]{Weaver77} Weaver, R., McCray, R., Castor, J., 
Shapiro, P. \& Moore, R., 1977, \apj, 218, 377

\bibitem[Weingartner \& Draine(2001a)]{Weingartner01a} Weingartner,
J.~C.~\& Draine, B.~T.\ 2001a, \apj, 548, 296

\bibitem[Weingartner \& Draine(2001b)]{Weingartner01b} Weingartner,
J.~C.~\& Draine, B.~T.\ 2001b, \apjs, 134, 263

%\bibitem[West(1976)]{West76}  West, R. 1976, \aap, 46, 327

%\bibitem[Whitmore et al.(1997)]{Whitmore97}  Whitmore, B.~C., Miller, B.~W.,
%Schweizer, F., \& Fall, S~ M. 1997, \aj, 114, 1797

%\bibitem[Wiklind, Combes, \& Henkel(1995)]{Wiklind95}  Wiklind, T., Combes,
%F., \& Henkel, C. 1995, \aap, 297 643

\bibitem[Wunderlich, Wielebinski, \& Klein(1987)]{Wunderlich87} 
Wunderlich, E., Wielebinski, R., \& Klein, U.\ 1987, \aaps, 69, 487

\bibitem[Yan et al.(1999)]{Yan99}  Yan, L., McCarthy, P.~J., Freudling, W.,
Teplitz, H.~J., Malumuth, E.~M., Weymann, R.~J., \& Malkan, M.~A. 1999,
\apj, 519, L47

%\bibitem[Young et al.(1995)]{Young95}  Young, S., Hough, Axon, D.~J.,
%Bailey, J.~A., \& Ward, M.~J. 1995, \mnras, 272, 513

\bibitem[Yun, Reddy \& Condon(2001)]{Yun01}  Yun, M.~S., Reddy, N.~A., \&
Condon, J.~J. 2001, \apj, 554, 803

%\bibitem[Zink et al.(2000)]{Zink00}  Zink, E.~C., Lester, D.~F., Doppman,
%G., Harvey, P.~M. 2000, \apjs, 131, 413

\end{thebibliography}
\end{document}